\documentclass[]{aa}
\usepackage{graphicx}
\usepackage[utf8]{inputenc}
\usepackage{color}

\begin{document}

\newcommand{\Tef}{$T_{\rm eff}$~}
\newcommand{\fel}{$f_{\rm el}$~}
\newcommand{\vsini}{$v\sin~i$~}
\newcommand{\lala}{$\lambda\lambda$~}
\newcommand{\Vt}{$V_t$~}
\newcommand{\DD}{$D_0$~}
\newcommand{\logg}{$\log$~g~}
\newcommand{\mum}{$\mu$m~}

\newcommand{\red}[1]{{\color{red} #1}}
\newcommand{\blue}[1]{{\color{blue} #1}}
\newcommand{\green}[1]{{\color{green} #1}}

\newcommand{\cm}{cm$^{-1}$}
\newcommand{\um}{$\mu$m}
\newcommand{\X}{X~$^3\Delta$}
\newcommand{\E}{E~$^3\Pi$}
\newcommand{\D}{D~$^3\Sigma^-$}
\newcommand{\A}{A~$^3\Phi$}
\newcommand{\B}{B~$^3\Pi$}
\newcommand{\C}{C~$^3\Delta$}
\newcommand{\F}{F~$^3\Sigma^+$}
\newcommand{\Sa}{a~$^1\Delta$}
\newcommand{\Sd}{d~$^1\Sigma^+$}
\newcommand{\Sb}{b~$^1\Pi$}
\newcommand{\Sg}{g~$^1\Gamma$}
\newcommand{\Sh}{h~$^1\Sigma^+$}
\newcommand{\Sc}{c~$^1\Phi$}
\newcommand{\Sf}{f~$^1\Delta$}
\newcommand{\Si}{i~$^1\Pi$}
\newcommand{\Se}{e~$^1\Sigma^+$}
\newcommand{\tsi}{$^{46}$Ti}
\newcommand{\tse}{$^{47}$Ti}
\newcommand{\tei}{$^{48}$Ti}
\newcommand{\tni}{$^{49}$Ti}
\newcommand{\tfi}{$^{50}$Ti}
\newcommand{\mc}{\multicolumn}
\newcommand{\TiO}{$^{48}$Ti$^{16}$O}
\newcommand{\LLname}{{\sc Toto}}

\newcommand{\Msol}{\ensuremath{M_{\odot}}}
\newcommand{\pp}{^{\prime\prime}}
\newcommand{\alert}[1]{\textcolor{red}{#1}}

\title{Analysis of the TiO isotopologues in stellar optical spectra}

\author{Ya. V. Pavlenko \inst{1,2} \and
Sergei N. Yurchenko \inst{3}
\and Laura K. McKemmish \inst{4}
  \and Jonathan Tennyson \inst{3}}

\institute{%
Main Astronomical Observatory of NAS Ukraine, 27 Zabolotnoho,  Kyiv, 01137, Ukraine
\and
Centre for Astrophysics Research, University of Hertfordshire, College Lane, Hatfield, AL10 9AB, United Kingdom
\and
Department of Physics and Astronomy, University College London, London WC1E 6BT, United Kingdom
\and
School of Chemistry, University of New South Wales, 2052 Sydney, Australia
}

\date{\today}

\abstract{
This study is based on the use of the new ExoMol TiO rovibronic line lists
to identify and model TiO isotopologue features in spectra of M-dwarfs.
}{
 We aim to investigate problems involving the computation of electronic bands for different isotopologues of TiO
by modelling optical spectra  of late-type stars  and to determine their Ti isotopic abundances,
to compare the TiO isotopologues spectra computed using line lists by different authors.
}{
We  fit theoretical synthetic spectra to the observed stellar
molecular bands of TiO. We model spectra of two M-dwarfs: GJ 15A (M1V) and GJ 15B (M3 V)
to determine Ti isotopic ratios in their atmospheres.
}{
We demonstrate the accuracy of the ExoMol \LLname{} line list for different
isotopologues of TiO and the possibility of determining  accurate Ti isotope abundances in a number
of spectral ranges. The 7580--7594 \AA\ spectral range seems particularly useful, with two atomic lines
of Fe I and molecular band heads  of \tfi\,O, \tni\,O, \tei\,O, \tse\,O{} clearly observable in our two M-dwarf spectra.
 We determine non-solar Ti isotopic ratios of  \tsi\ : \tse\ : \tei\ : \tni\ : \tfi  = 7.9 : 5.2 : 72.8 : 7.9 :6.2  for GJ 15A and
7.4 : 4.2 : 76.6 : 5.8 : 6.0 for GJ 15B with accuracy of order $\pm$  0.2. [Ti] = 0.040 and 0.199 within accuracy $\pm$ 0.10 were also
determined for GJ 15A and GJ 15B, respectively.
}{
 We find that the ExoMol \LLname{} TiO line list:
a) describes the fine details in line position and intensities of the M-dwarf spectra better than other available TiO line lists;
b) correctly reproduces the positions and intensities of the TiO isotopologue band heads  observed in M-dwarf spectra;
c) can be used to determine Ti isotope abundances in atmospheres of M-stars.
}

\maketitle

\titlerunning{Analysis of the TiO spectra}
\authorrunning{Ya. Pavlenko et. al.}


\section{Introduction}

Naturally occurring titanium, the element of nuclear charge 22, is composed of five stable isotopes;
\tsi, \tse, \tei, \tni ~and \tfi. Thus far,  nuclide abundances of these elements in the solar system formed 4.56 Ga ago
have been measured with high accuracy. \tei\ is the most abundant isotope (73.72\% natural abundance) with solar system isotopic ratios of
\tsi : \tse : \tei : \tni : \tfi= 8.249 : 7.372 : 73.72 : 5.409 : 5.185 \citep{lodd09},
similar numbers are given by \cite{biev93}.  Atoms of Ti isotopes in low temperature regime (T < 4000 K) are bound in TiO molecules, molecular bands of TiO
are used to assess the low-mass stars and brown dwarfs of spectral class M.
Twenty-one radioisotopes of Ti have been also characterized, with the most stable being $^{44}$Ti with a half-life of 60 years.
  $^{44}$Ti is observed in the supernova (SN) remnants, see \cite{moch99}. This isotope  offers  one of the most
direct probes into nucleosynthesis environments in the
interior of exploding stars, when the associated gamma-ray
activities in the explosion remnant are detected and
translated back to the isotopic abundances at the time of the explosion
\citep{moch04, moch08}.

Details  of the galaxy nucleosynthesis of Ti isotopes are described in many papers, e.g. see \cite{hugh08}.
 Sources of the stable Ti isotopes are known
pretty well, see table 1 in \cite{alex99} as well as \cite{woos94}:

-- \tsi~ and \tse~ are formed during explosive oxygen and silicon burning, respectively, in  SN
Types II and Ia;

-- \tei~ is formed by explosive Si burning in  Type II SN;

-- \tni~ is formed by the explosive Si burning in Type II SN;

-- \tfi~ is formed in nuclear burning in  Type Ia SN.

\cite{moch99} noted,
 that long-lived isotopes with mean lifetimes of the
order of 10$^6$ yrs or longer will reflect a superposition of different
supernovae at different times, mixed with interstellar matter.
Consequently, abundances in an individual object are
formed by different branches of many processes at large time scales.
Only short-lived isotopes will clearly trace
individual events.

The classical model of \cite{timm95} suggested that most stable Ti isotopes
were under-produced with respect to solar values: relative to \tei\ and
the solar system isotopic ratios, the predicted abundances of  \tsi, \tse,
\tni, and \tfi\, are factors of 2 too large, 3 too small, spot on,
and 2 too small, respectively. Isotopic ratios were predicted to
decline with decreasing [Fe/H] by factors of 8 for \tsi, 6 for
\tse, 2 for \tni, and 30 for \tfi\, between [Fe/H]= 0 and -1.
The degree of this underproduction depends on many theoretical assumptions, see \cite{hugh08}.

Essentially, modelling the abundances of TiO isotopologues \citep{lamb72,wyck72,lamb77,cleg79,vale98}  is the only way to 
extract Ti isotopic abundances from M-stars. It is a very hard task to determine Ti isotopic ratios from analysis of atomic lines due to small isotopic shifts of different isotopic lines, see \cite{koba19} and references therein. Indeed, broadening of spectral 
lines by macroturbulence in the stellar atmospheres as well as any notable rotation of stars makes separating
of the atomic spectra of different isotopes almost impossible \citep{teny19}.

Accurate molecular line lists  - i.e. lists of energy levels and the intensity of spectral transitions - are crucial for correctly modelling a
variety of astrophysical phenomena, including stellar
photospheres and the atmospheres of extra-solar planets.
Transition metal diatomic line lists are particularly difficult due to the large
number of coupled low-lying electronic states  \citep{jt632}.
TiO line lists are particularly important for studies of M-dwarfs
\citep{pavl95, alla00} and a number of them have been constructed \citep{schw98,plez98,ryab15,jt760}.
The most recent of these is the \LLname{} line list of \cite{jt760} which was computed
as part of the ExoMol project \citep{jt528}.

TiO line lists were first used to extract titanium abundance in the 1970s, e.g. by \cite{lamb72,wyck72,lamb77,cleg79}. These studies were limited, however, by the quality of the underlying line list, leading \cite{vale98}  to introduce isotope-dependent shifts to the line positions. These modifications should be no longer necessary with modern line list development techniques.

 The majority of  previous studies used the spectral range 7053.5 - 7055 \AA~ to quantify abundance of TiO isotopologues and
thus Ti abundance. However, the quality and availability of new more accurate TiO molecular spectroscopic data and
analysis techniques means the suitability of this spectral region should be reconsidered.

 We note that detailed comparisons with  high resolution M-dwarf spectra
performed by \cite{jt760} showed that their \LLname{} line list performed much better than
the earlier line lists, particularly in matching line positions. These comparisons
naturally concentrated on the main, \tei, isotopologue. In this paper we assess
the utility of the \LLname{} line list for modelling TiO isotopologue spectra in M-dwarfs.
A more detailed discussion of the \LLname{} line list is given below.
We also extended the comparison of the quality of TiO isotopologue
line lists computed by different authors, see Section  \ref{_ltio} using the comparison of fits of computed spectra
to the observed spectra.

The nearest and brightest stars are among the best studied and
hold a special place in the popular imagination. The discovery of
planets orbiting these stars tells us that the solar neighborhood
is potentially rich with exoplanet systems, see \cite{howa14} for
the complete list of known exoplanet-hosting stars within 7 pc. The big advantage of
investigating  these stars is that it is  possible
to undertake extensive campaigns to characterize them using a
combination of high-resolution optical spectroscopy, near infrared (IR)
spectroscopy, long-baseline optical/infrared interferometry, and
high-cadence, broadband optical photometry.
 Even for  the history of our own planetary system the Ti isotopic studies are
of great importance. \cite{zhan12} found that the \tfi/\tse~ ratio of the Moon is
identical to that of the Earth within about four parts per
million, which is only 1/150 of the isotopic range documented in
meteorites, see table 2 in \cite{zhan12}.
It means that the Moon-forming giant impactor, i.e hypothetical Theria,
would be of the same isotopic composition as our Earth.
Indeed, theory predicts that the Moon ends up constructed mostly (40 -- 75\%)
from the impactor materials.
However,  the observed identical titanium (and oxygen) isotopic compositions in the
Earth and the Moon are surprising in light of what we think we know about
planet formation and formation of the Moon after a giant impact, see \cite{tayl12}.

We expect growing interest of exoplanetary astrophysicists to
isotopic studies of Ti and other elements.

In this paper we study the TiO isotopologues in spectra of the wide binary system GJ 15 A,B.
GJ 15A (also known as Groombridge 34A, V* GX And, HD 1326A, HIP 1475, BD+43 44A,
Gaia DR2 385334230892516480) is a cool red dwarf of type
M1. The other member of
this binary star system, GJ 15B (also known as
Groombridge 34B, V* GQ And,  HD 1326B, BD+43 44B, Gaia DR2 385334196532776576) is fainter and has a spectral type
of a M3.5 V \citep{reid95}. \cite{lipp72} measured
a small astrometric segment of their orbit, giving an AB
separation of 146 AU and an orbital period of 2600 yr.
Based on an imaging search for companions at 10 \mum with
MERLIN at Palomar, \cite{bure98} ruled out additional
companions to A having projected separations of 9–36 AU with
\Tef $>$ 1800 K (M $>$ 0.084 M$_{\odot}$). \cite{gaut07} found no
infrared excess for GJ 15A at 24, 70, or 160 \um.
Since GJ 15A is a bright nearby star, many teams have provided
studies of this component of the binary system; these are summarized in
Table \ref{_gj15info}.

\begin{table*}
\caption{\label{_gj15info} Parameters of stars taken from the literature.}
\begin{tabular}{lllll}
\hline
\hline
\noalign{\smallskip}
\Tef   &  \logg & [Fe/H] & CompStar &      Reference   \\
\noalign{\smallskip}
\hline
\noalign{\smallskip}

\multicolumn{5}{c}{GJ 15A} \\
\hline
\noalign{\smallskip}
  3606  & 4.93    & -0.27   &        & \cite{2018AA...615A...6P}  \\
  3669  &         & -0.29   & SUN    & \cite{2014MNRAS.443.2561G} \\
  3693  &         & -0.26   & SUN    &\cite{2014ApJ...791...54G} \\
  3551  &         &-0.28    &        &\cite{2015ApJS..220...16T} \\
  3760  &         &-0.23    &        &\cite{2015ApJS..220...16T} \\
  3368  &         &-0.26    &        &\cite{2015ApJS..220...16T} \\
  3603  &         & -0.30   & SUN    &\cite{2015ApJ...804...64M} \\
  3988  &         &         &        &\cite{gaia18}              \\
\hline
\noalign{\smallskip}
\multicolumn{5}{c}{GJ 15B} \\
\hline
\noalign{\smallskip}
  3283    & 5.11 &-0.19   &         &  \cite{2018AA...615A...6P}  \\
  3282    &      &-0.17   &SUN      &  \cite{2014MNRAS.443.2561G} \\
  3540    & 5.30 &-0.20   &SUN      &  \cite{1998MNRAS.299..753Z} \\
  3254    &      &-0.20   &SUN      &  \cite{2014ApJ...791...54G} \\
  3218    &      &-0.30   &SUN      &  \cite{2015ApJ...804...64M} \\
          &      &-0.08   &SUN      &  \cite{2014AJ....147...20N} \\
  3679    & 4.92 &-1.15   &SUN      &  \cite{2011AA...531A.165P} \\
  3630    & 4.71 &-0.88   &SUN      &  \cite{2012AA...538A.143K} \\
  3330    & 5.08 &-1.40   &SUN      &  \cite{2007MNRAS.374..664C} \\
  3636    &      &        &         &  \cite{gaia18}              \\
\hline\hline
\end{tabular}
\end{table*}

\section{Procedure}

In this paper we model optical spectra of M-dwarfs. TiO features dominates here, however, other
molecules also provide notable features in different spectral ranges, see \cite{pavl14}\footnote{Color plots are available at
www.mao.kiev.ua/staff/yp/Results/M-stars/mb.htm}.
Besides TiO, absorption by molecular bands of VO, CaH, CrH, MgH and other
hydrides from Kurucz database \citep{kuru11}
were accounted in our computations, as well as absorption by atomic lines taken from VALD \citep{ryab15}.

\subsection{Titanium oxide line lists}
\label{_ltio}

In the following, we analyse different line lists for monosubstituted
isotopologues of TiO computed using the \LLname{} line lists of \cite{jt760}. The
\LLname{} line lists were constructed for the main isotopologues of titanium oxide (TiO), namely {\tsi}O, {\tse}O, \tei\,O,
{\tni}O and {\tfi}O. Here and below we assume that oxygen
is represented by its $^{16}$O isotope as other oxygen isotopes are much less abundant in both the solar vicinity
\citep{biev93,lodd09} and the majority of  the known astrophysical objects.

The \LLname{} line lists contain transitions with wavenumbers up to
30,000 cm$^{-1}$, i.e. long-wards of 330 nm, and include all dipole-allowed transitions
between  13 low-lying electronic states (\X, \Sa, \Sd, \E, \A, \B, \C, \Sb, \Sc, \Sf, \Se).
The \LLname{}  rovibronic line positrons were constructed using potential energy curves as simple Morse oscillators with constant diagonal and off-diagonal spin-orbit and other coupling terms fitted to match known empirical energy levels and {\it ab initio} curves where experimental data was unavailable. Accurate line intensities were generated using {\it ab initio} dipole moment curves.
Final line lists were computed using the variational nuclear-motion program {\sc Duo} \citep{jt609} where
various couplings were explicitly included.
The \LLname{} line lists are appropriate for temperatures below 5000 K and contain about 60 million
transitions for the main TiO isotopologue; higher temperature data is not required as TiO bands disappear in stellar spectra above about 4200 K due to dissociation of the titanium oxide molecule.

The variational procedure described above is known not  to provide sufficiently accurate TiO spectra
for high resolution studies, see \cite{15HoDeSn.TiO} for example. Therefore, \cite{jt760}
used empirical energy levels they had obtained previously \citep{jt672} using the so-called
MARVEL (measured active rotation-vibration energy levels) procedure \citep{jt412,jt750} to
improve the \LLname{} energy levels and thus predicted transition frequencies. MARVEL energies are only available for \TiO; the comparisons made by \cite{jt760} show that these corrected frequencies reproduce
the majority of stellar features associated with TiO very well.

For the other isotopologues \cite{jt760} adopted the method of \cite{jt665} and shifted the isotopologue
energy levels by the observed (MARVEL) minus calculated value obtained for \tei O. They made no
allowance for any shifts associated with breakdown of the Born-Oppenheimer approximation.
\cite{jt760} provides initial evidence for the success of this approach in  the region 14145 - 14175 cm$^{-1}$ by comparing
 experimental TiO spectra to cross-sections generated using the \LLname{} line list and demonstrates that the weak peaks  in the experimental spectra are isotopologue peaks.
The present study provides a stringent test of how well this method works in practice.

The  \LLname{} TiO spectroscopic data  is given in two datasets, a states and a transitions file \citep{jt631}.
The huge amounts of the data are not easy to use directly in
astrophysical computations. Therefore we
converted the   \LLname{} line data to the format of our synthetic spectra computation  programs,
see Table 1 in Appendix.
Furthermore,
to reduce the number of lines which could effectively absorb
radiation in lower temperature regime considered here we select only stronger lines
using the cutoff parameter $\alpha > 10{-6}$, here
\begin{equation}
 \alpha = gf * \exp (-E''/kT),
\end{equation}
where $f$ is the oscillator strength of line in absorption, $g$ is the statistical weight of lower level,
$E''$ is the lower state energy, $T$= 3000 K.
After computing the wavelength of the lines
in vacuum they were shifted to the air wavelength scale using formulae of
\cite{cidd96}. In the following we use the air wavelength scale.

 To estimate the quality of the different line lists, we compared fits to the
observed spectra of GJ 15A and GJ 15B using line lists of
TiO isotopologues computed by \\
-- \cite{schw98}, labelled S98 below, \\
-- \cite{plez98}, an updated version was  taken from the web page\footnote{https://nextcloud.lupm.in2p3.fr/s/r8pXijD39YLzw5T?path= \\
\%2FTiOVALD}, labelled P12 below,\\
--  \cite{jt760}, the \LLname{} line list.

Results of our comparison of synthetic spectra computed with different lists with observed spectra are given in Section \ref{_xx1} and \ref{_xx2}.

\subsection{Observed spectra}

For the analysis, we used high resolution spectra of both stars from the
CARMENES spectral library of \cite{rein18}.
 The CARMENES spectra cover the wavelength range 520 -- 1710 nm at a resolution of at least R = 80,000 or better.
The spectra of GJ 15A and GJ 15B used in this work, were observed with exposures in 61 and 198~s to get
 the comparable S/N = 68 and 54, respectively. We refer to \cite{rein18}  for more details.
All theoretical spectra were shifted to account for the radial velocity of the appropriate star.

\subsection{Synthetic spectra, model atmospheres}

To generate synthetic spectra we used the BT-Settl model atmosphere \citep{alla14}. Procedure for computation of synthetic spectra are
described elsewhere, see \cite{pavl97}. We adopted the solar abundances of \cite{ande89}. Line profiles were
computed using Voigt profiles, damping constants were taken from the line list databases or computed
in the framework of the Unsold approximation. Synthetic spectra  were computed in
wavelength steps of 0.025 \AA, where we adopted the microturbulent velocity of \Vt = 1 km/s. Some numerical experiments showed
rather marginal dependence of our results on $V_t$. Our fitting procedure did not reveal any notable rotational velocities
$v\sin~i$, therefore the theoretical spectra were convolved with a pure Gaussian profile in order to model the instrumental broadening, see next subsection.  We fixed \Tef = 3800 K and \Tef = 3500 K for the A and B components, respectively. These values are not far from the Gaia's 3869 and 3636 K \citep{gaia18}, respectively.

\subsection{Fits to the observed spectra}
\label {_fos}

The best fit to the observed spectra was achieved  by the $\chi^2$ procedure described elsewhere \citep{pavl14}.
We give a few details here to aid understanding of our procedure. As part of the fit, the function
\begin{equation}
S= \sum_{i=1}^{N} s_i^2
\end{equation}
is minimised, where $s_i= |F^{\rm obs}_i - F^{\rm comp}_i|$; $F^{\rm obs}_i$ and $F^{\rm comp}_i$ are the observed
and computed fluxes, respectively, and $N$ is the number of the wavelengths points  used in the minimisation procedure.
Three parameters used in our minimisation procedure, i.e. the Doppler shift measured in km/s, the flux scale
normalisation parameter and the FWHM used for the smoothing Gaussian, were determined for every fitted spectral range.
In our analysis we omit some spectral
ranges which contain artifacts provided by strong noise, telluric absorption, bad pixels, etc.
The minimisation sum $S$ is  computed on a 3D grid of radial velocity sets, flux
normalisation factors, and FWHM parameters. Errors in the fit are evaluated as $\Delta S = \sum s_i/N$.

\section{Results}

\begin{table*}
\caption{\label{_tt} Spectral ranges of our interest.}
\begin{tabular}{lccllll}
\hline
\hline
\noalign{\smallskip}
Sp. range   & Wavelengths (\AA), air & Wavenumber (\cm{}) & El. system & ($v''$,$v'$)  \\
\noalign{\smallskip}
\hline
\noalign{\smallskip}
  $	x_1$  & 7053.5 -- 7055 & 14173.2 - 14170.2  & $\gamma$  (\A{} - \X{}) & (0,0)     \\
  $	x_2$  & 7580 -- 7594   & 13188.8 - 13164.5  & $\gamma$ (\A{} - \X{}) & (0,1)    \\
  $	x_3$  & 8194 -- 8204   & 12200.5 - 12185.6  & $\delta$ (\Sb{} - \Sa{}) & (1,0)   \\
  $	x_4$  & 8858 -- 8862   & 11286.0 - 11280.9  & $\delta$ (\Sb{} - \Sa{}) &(0,0)     \\
  $	x_5$  & 9720 -- 9737   & 10285.1 - 10267.1  & $\delta$  (\Sb{} - \Sa{})  &(0,1)   \\

\noalign{\smallskip}
\hline
\end{tabular}
\end{table*}

\subsection{Search for the best spectral ranges}

 Spectral features associated with isotopologues are difficult to see because they are faint
and close to each other. The flux ratio method shows the strongest TiO absorption regions that
should be analyzed for the largest splitting. This should be complemented by observations
of highest spectral resolution ($>$100,000) and highest S/N, see \cite{vale98}. It it worth noting here that
usually the presence of macroturbulence in the atmospheres of late-type stars reduces the effective resolution down to
70,000 or less.
Generally, the use of flux ratios could provide an alternative approach but
restricting ranges to the band heads allows the procedure to focus on
measuring isotopic ratios. It is possible that the use of  much wider
spectral ranges may offer reasonable statistics on the
accuracy of the individual wavelengths in the new line list  of the better quality.

 On the other hand, going to the shorter
wavelengths may also enhance the chances of finding high quality
high-resolution spectra in observational archives
(HIRES, HARPS etc.), see \cite{vale98} and others referenced in the introduction.
However, this approach requires  very high accuracy of the input line lists
and even minimum blending with other molecular lines, even from the same molecule, is likely to cause
problems.  In practice the TiO optical spectrum comprises absorption
bands of different systems and the responses of these band systems
to changes in the temperature structure of the model atmosphere are different due the differences in  excitation of the lower level of
each transition. Furthermore, using small spectral ranges is better than using cross-correlation across a large spectral range, and TiO the line list is less accurate at shorter wavelengths than longer wavelengths due to absence of reliable spectroscopic data for the molecular levels with high excitation, see \cite{jt760}
Conversely, we know from experience that analysis of the well-characterised isotopologue band heads across comparatively short spectral ranges
allows increasing accuracy of abundance determination by modelling  infrared molecular bands of CO and SiO, see \cite{pavl20} and references therein.

\begin{figure*}[h]
   \centering
   \includegraphics[width=0.98\columnwidth]{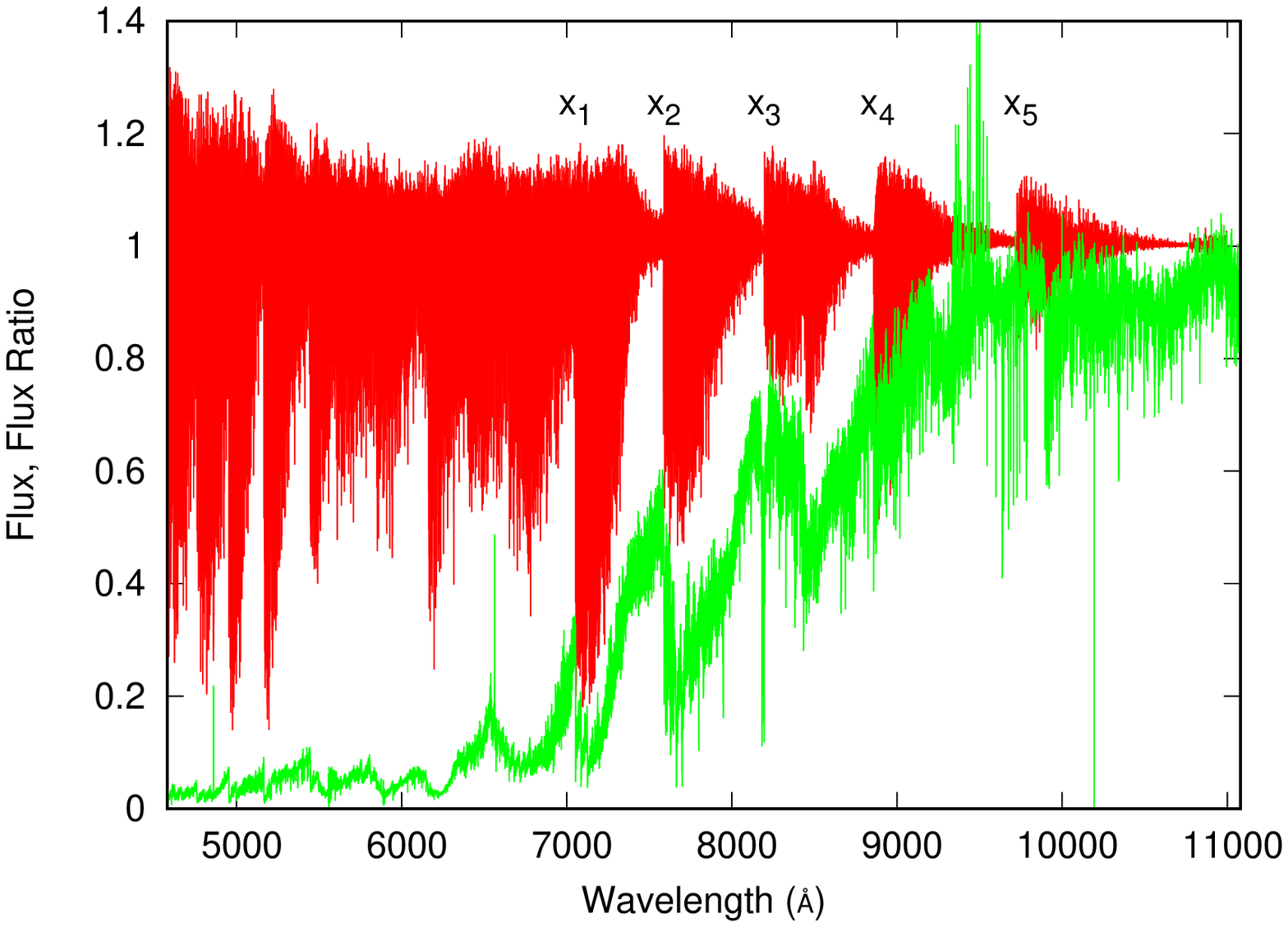}
    \includegraphics[width=0.98\columnwidth]{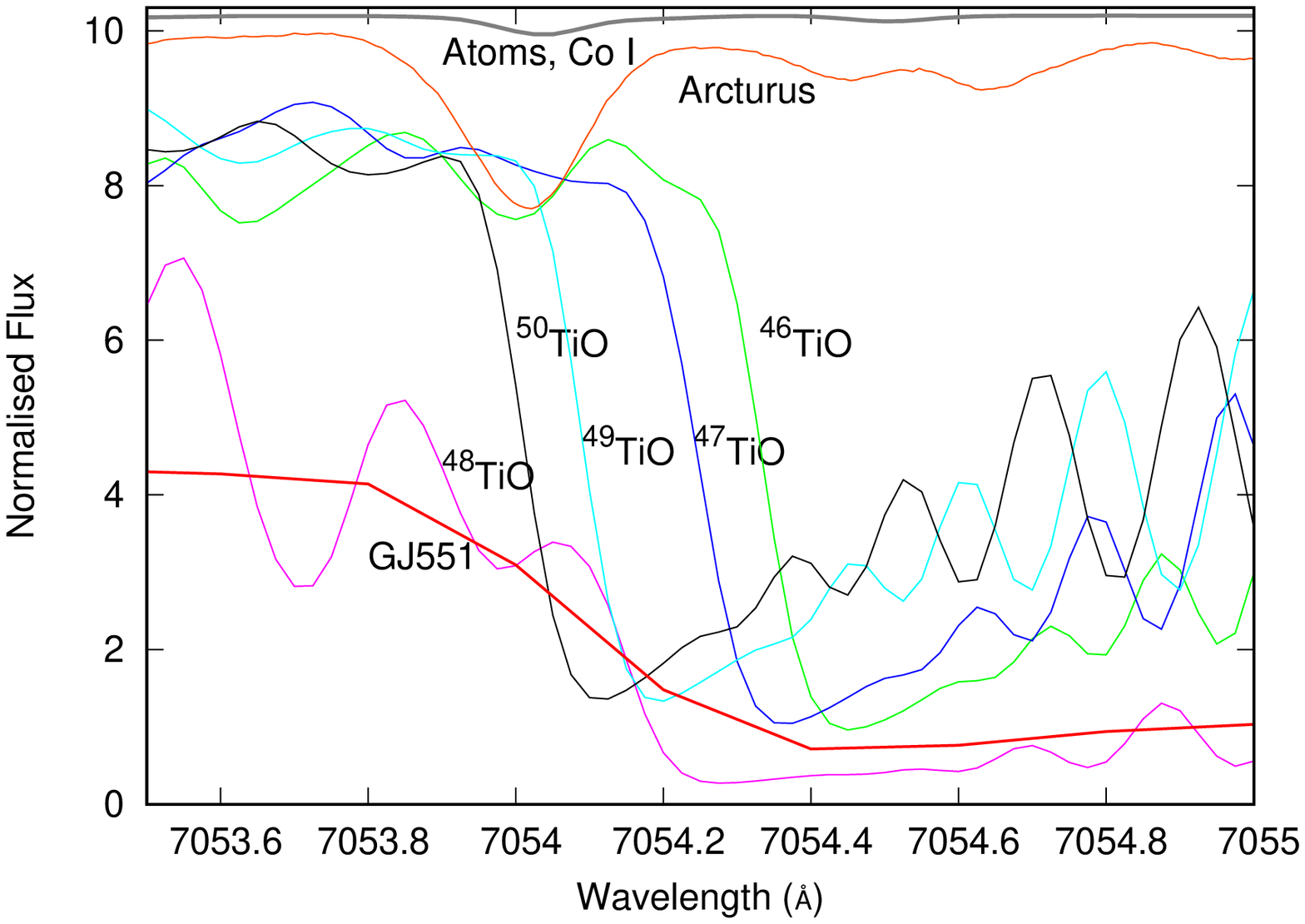}
    \includegraphics[width=0.98\columnwidth]{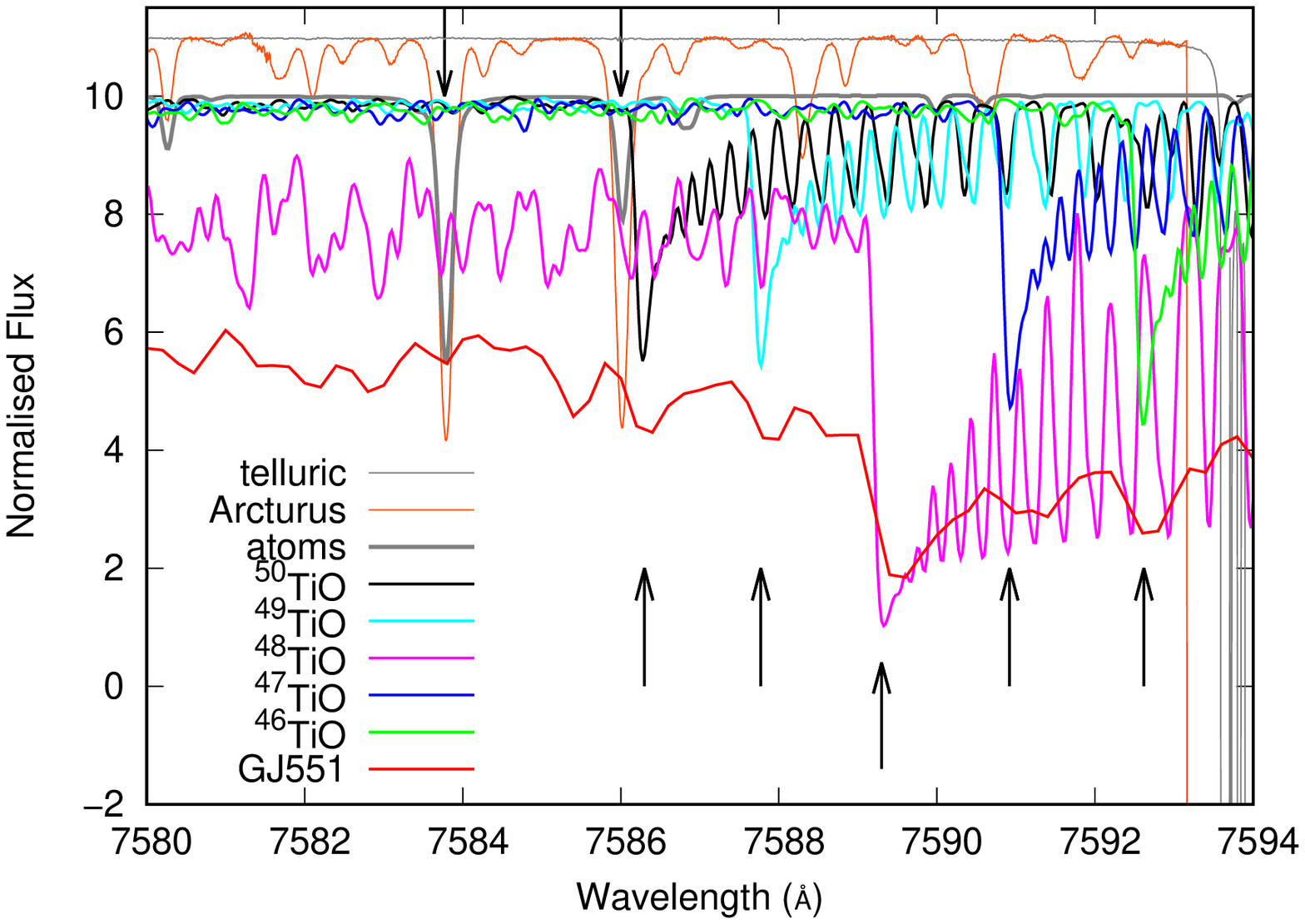}
    \includegraphics[width=0.98\columnwidth]{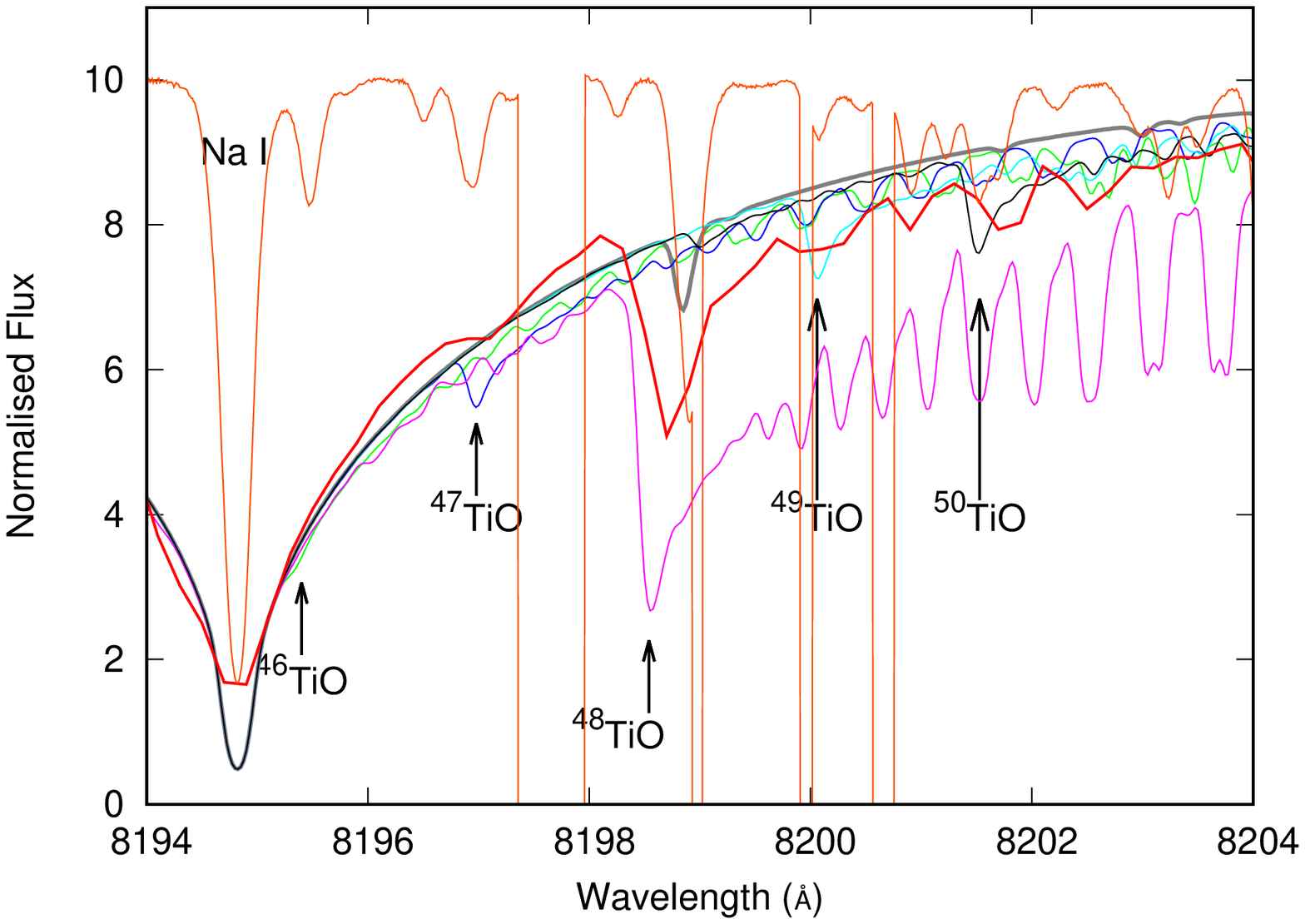}
    \includegraphics[width=0.98\columnwidth]{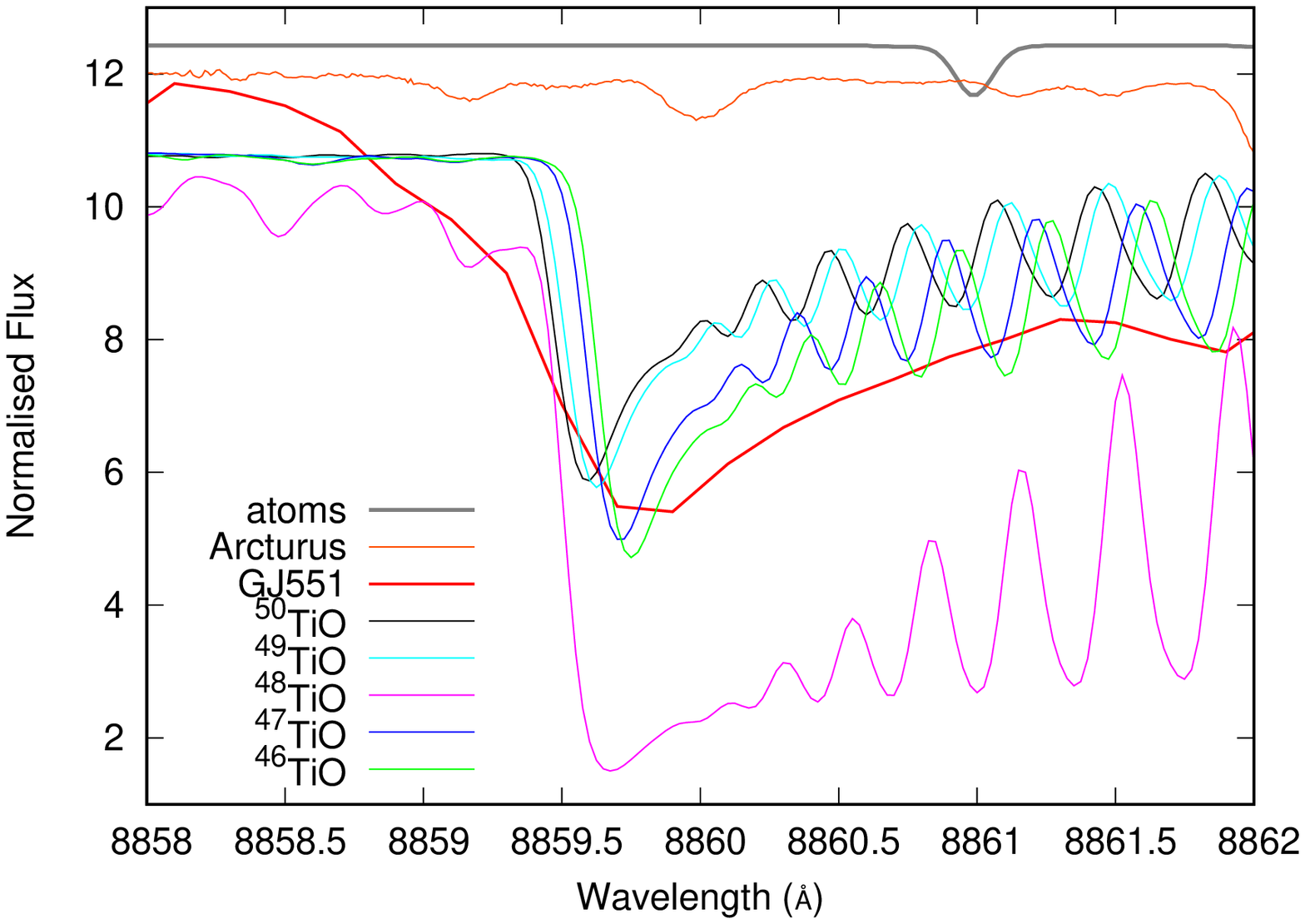}
    \includegraphics[width=0.98\columnwidth]{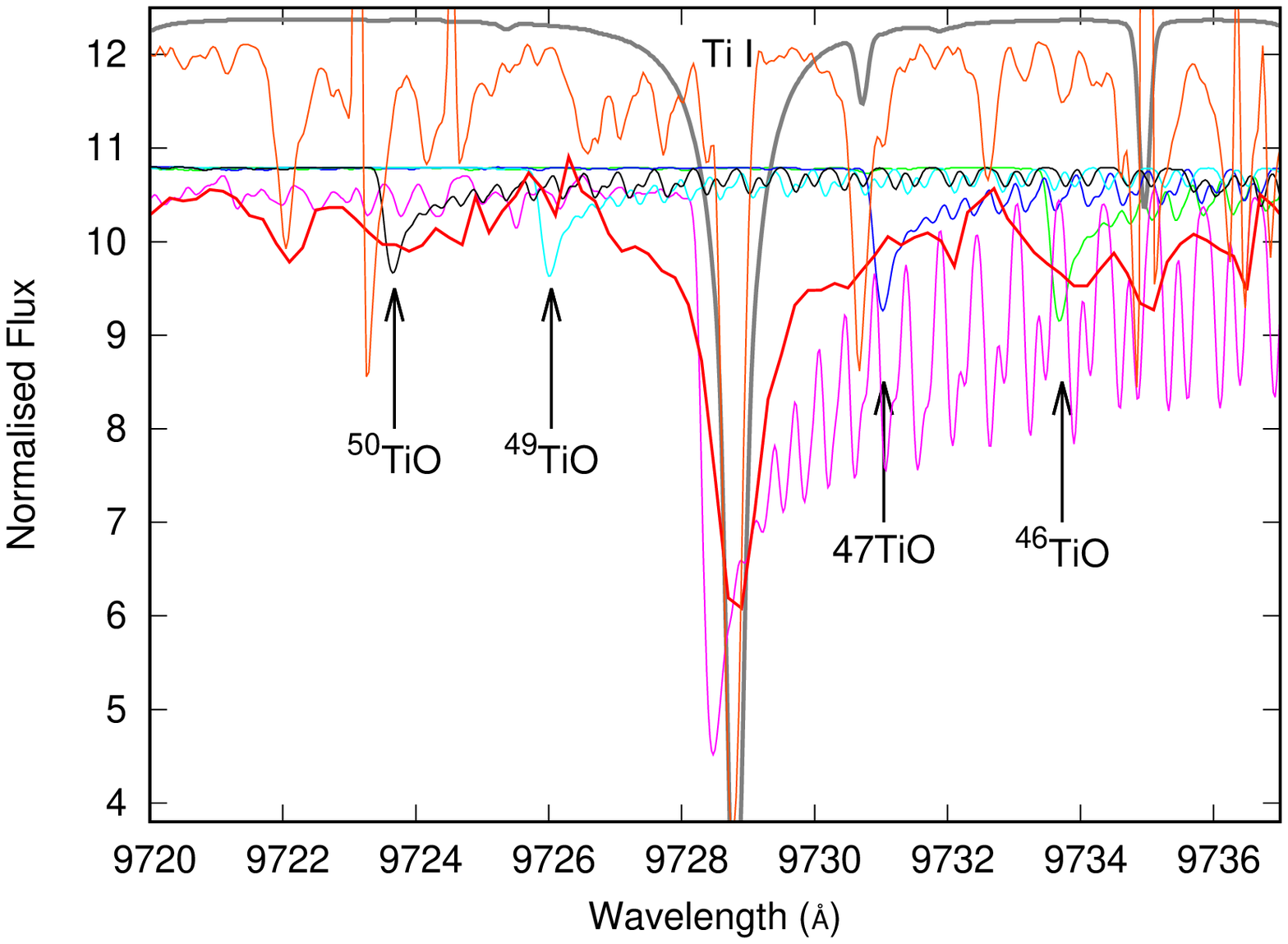}
\caption{\label{_2945} {\it Top left:} Observed X-shooter spectrum of GJ551 (green line) \citep{pavl17} and flux ratio
$F_{48}/F_{all}$ (red line); the green line shows the observed X-shooter fluxes in the Proxima spectrum.
$x_1, ... x_5$ mark our spectral ranges of interest.
{\it Top right:} Band heads of TiO isotopologues computed for a 2900/4.5/0.0 model atmosphere.
All observed spectra are shifted to the zero radial velocity frame.
{\it Middle panels:} Theoretical spectra of bands of TiO isotopologues computed
across the $x_2$ {it (right panel}) and $x_3$ {\it (left panel)} spectral ranges. Upward arrows show the positions of
head bands of the TiO isotopologues.
Downwards arrows label two Fe I lines seen in absorption.
{\it Bottom panels:} Theoretical spectra of bands of TiO isotopologues computed
across $x_4$ {\it (right panel)} and $x_5$ {\it (left panel)} spectral ranges.
Position of the molecular band heads are shown by arrows. In all panels
observed spectra of Acturus and Proxima Cen (GJ551) are shown by thin and thick red lines, respectively, to
demonstrate locations of notable atomic and molecular features in the observed spectra.
}
\end{figure*}

Therefore, as a first step in our analysis, we determine which spectral regions are useful for Ti isotopic abundance determination. To this end we computed two
spectra of fluxes: the first spectrum, $F_{48}$, contains absorption of {\tei}O only, while the second spectrum $F_{all}$  contains all TiO isotopologues assuming the ``solar'' Ti isotopic ratios.
Both spectra were smoothed by Gaussians with $R=60,000$ and are shown in Fig. \ref{_2945} together with their Flux ratios: $F_r = F_{all}/F_{48}$.
The flux ratios across band heads should show the largest sensitivity to the isotopologue abundances.
Here we consider several spectral ranges
with $\lambda > 7000$ \AA. At shorter wavelengths, spectra of M stars become more complicated due to interference between
different molecular bands while their flux  also drops noticeably  bluewards.
We denote the spectral ranges of interests as $x_1, x_2 \ldots x_5$ (see Table \ref{_tt}) and analyse the computed flux ratios in comparison with the observed X-shooter
spectrum of GJ551 (Proxima Cen) as described by \citet{pavl17}.

\subsubsection{Spectral range $x_1$: 7053.5 -- 7055 \AA}

The main spectral feature of $x_1$ is formed by the TiO $\gamma$-system, i.e. \A\ -- \X, (0,0) band with its leading red-degraded band head
at 7054 \AA. This spectral window was used for modelling sunspot spectra by \cite{lamb72},
Omicron Ceti by \cite{wyck72}, Aldebaran by \cite{lamb77}, spectra of late-type dwarfs and
giants by \cite{cleg79}, studies of M-dwarfs by \cite{vale98} and spectra of local M-dwarfs by \cite{chav09}. Unfortunately, the TiO band head in this region
is formed by blends of unresolved components of TiO isotopologues, see the top right panel of Fig. \ref{_2945}, therefore
we preferred to use wavelengths longer than the
molecular head at 7054 \AA.
Furthermore, at the wavelength of the band head there is a strong atomic line, clearly seen in the spectrum of Arcturus \citep{hink95},
which further complicates the analysis.

It is worth noting that this spectral range is also affected by absorption due to strong lines of the TiO $\gamma'$ (0,1)
band which is a part of the \B\ -- \X\ band system the
(0,0) band head at 6192.5 \AA. A mixture of lines of the $\gamma$ and $\gamma'$ bands complicates the fine structure
analysis of the $x_1$ spectral range. Note both bands are included in the \LLname{} line list.

\subsubsection{Spectral range $x_2$: 7580 -- 7594 \AA}
\label{_xxx1}

 The head of the red-degraded (0,1) band of the $\gamma$-system (\A\ -- \X{}) is located at 7589 \AA.  Here we have some
contribution from lines in the tail of the (0,0) TiO band, which becomes stronger at lower \Tef.

Generally speaking the fluxes in the optical spectra of late type dwarfs increase to the red,
therefore this spectral range is preferable. On the other hand, the strong telluric O$_2$ A-band
absorption is located
approximately at $\lambda = $ 7580 \AA\  \citep{rudo16}. We note that the position of the molecular head in stellar spectra reduced to the local coordinate frame  varies depending on the relative radial velocity of the star.
In the X-shooter spectrum of Proxima Cen obtained with long exposures,
the O$_2$ molecular band is found at even  shorter wavelengths, i.e. beyond 7580 \AA.
There are two atomic lines of Fe I in the $x_2$ spectral range  which provide the means
to determine accurate Doppler shifts important for the proper
identification of the weak TiO isotopologue features.

As shown in the left middle panel of Figure \ref{_2945}, features created by
{\tfi}O, {\tni}O, and, of course, {\tei}O
are clearly seen in the computed HIRES spectra  as well as in the X-shooter spectrum of intermediate resolution of Proxima. The heads of the of isotopologues bands
{\tsi}O and {\tse}O can be seen only in the background of stronger
lines belonging to the tail of {\tei}O in the HIRES spectra.

An additional advantage of the $x_2$ spectral region is the presence of two atomic Fe I lines at 7583.8 and 7586.0 \AA,
see the  downward arrows in the middle left
panel of Fig. \ref{_2945}. These lines are clearly seen in the Arcturus spectrum, and  can be used for accurate wavelength reduction in cooler stars,
as well.

\subsubsection{Spectral range $x_3$: 8194 -- 8204 \AA}

The $x_3$ spectral range contains the (1,0) band head of the $\delta$ (\Sb\ -- \Sa) system of TiO.  The main absorption
feature here is formed  by the reddest line of the subordinate Na I triplet at 8183.255, 8194.790, and 8194.823
 \AA (8200 \AA\ triplet), see the right middle panel of Fig.~\ref{_2945}.
 Interestingly, the molecular bands occur in the reverse order in comparison to the $\gamma$ band system;  here the heads of the isotopologues bands occur in ascending order, i.e. from {\tse}O to {\tfi}O.

Unfortunately, that structure of the $x_3$ spectral range makes the detailed analysis of the weak molecular features created by the
bands of TiO isotopologues difficult. Subordinate Na I lines  of the 8200 \AA~ triplet
are well known indicators of the gravity in the late type dwarfs, but they depend on the
effective temperature as well, see \cite{schl12}. Likely, the $x_3$ wavelength range can be used for Ti
isotopic analysis
in spectra of the lower gravity stars, in which Na I lines at 8200 \AA~ should have weaker wings.

\subsubsection{Spectral range $x_4$: 8858 -- 8862 \AA}

The strong head of the (0,0) band of the $\delta$ system, i.e. \Sb\ -- \Sa, is a notable observational feature in
the $x_4$ range. Unfortunately, the wavelength shifts between heads of TiO isotopologue bands are small so that even in the HIRES spectra one can see only one band head formed by several TiO isotopologues bands. The tail of the composite molecular band provides some opportunities to detect isotopologue lines which are notably shifted with respect to the {\tei}O lines; these can be seen in the 8881 -- 8862 \AA\ range where strong atomic or telluric
lines are absent. On the other hand, these wavelengths are severely polluted by (0,0) FeH and CrH molecular lines of the
FeH F~$^4\Delta_i-$ X~$^4\Delta_i$ and CrH A~$^6\Sigma^+-$ X~$^6\Sigma^+$ systems with band heads at 8694 and 8520 \AA, respectively, see
\cite{pavl14}.

\subsubsection{Spectral range $x_5$: 9720 -- 9737 \AA}

The $x_5$ spectral range contains the  (0,1) band head of the $\delta$ (\Sb\ -- \Sa) system of TiO.
The bottom right panel of Fig. \ref{_2945} illustrates at least two problems of this wavelength
range: (a) the spectrum is severely polluted by telluric absorption; (b) the strong Ti I atomic line at 9728.32 \AA\ seen in the spectrum of Arcturus and Proxima forms blend with a
molecular band head of {\tei}O.
On the other hand, the  TiO isotopologue band heads are well spaced in wavelength.
Again, molecular bands of  heavier isotopologues are shifted blue-ward, away from the strong  {\tei}O band head.
The $x_5$ spectral range can be used for determination of {\tfi}O and \tni O abundances but only in the
of absence of telluric absorption at the wavelengths of their band heads.
The {\tse}O and \tsi O band heads can be observed in the background of
the stronger tail of the {\tei}O band. Again, the strong telluric absorption makes  it difficult to
perform fine analysis
of the weak molecular features across this spectral range.

\subsection{Determination of Ti isotope abundances in atmospheres of GJ 15A and GJ 15B}

After selection of the best spectral ranges for the determination of Ti isotope abundances, we
performed fits of our synthetic spectra to observed spectra of GJ 15A and
GJ 15B across selected spectral ranges to demonstrate the possibility of using them to determine abundances of Ti isotopes in their atmospheres.
As before we assumed that all oxygen atoms exist in the form $^{16}$O.

\subsubsection{Spectral range $x_1$}
\label{_xx1}

\begin{figure*}
    \includegraphics[width=0.98\columnwidth]{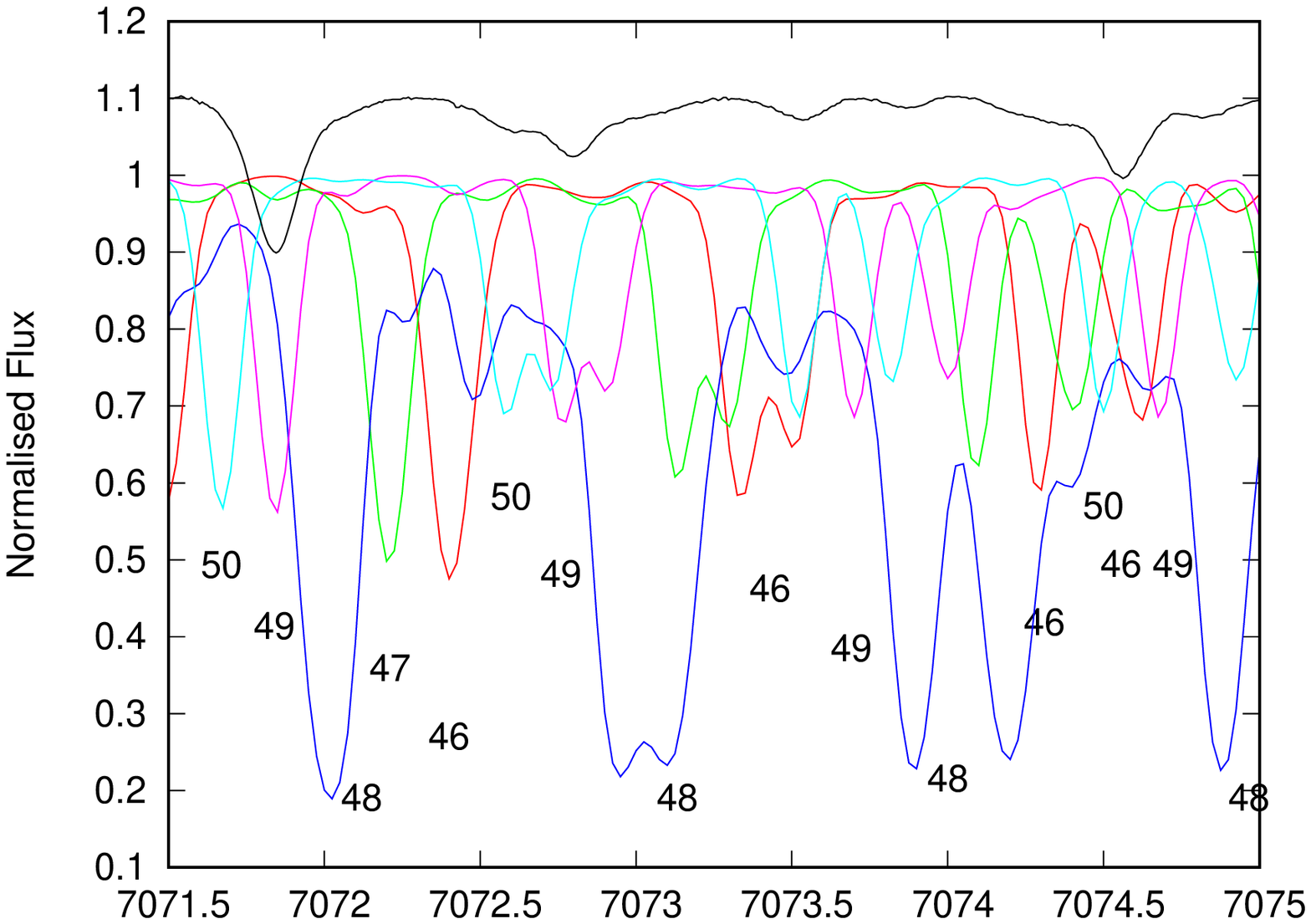}
    \includegraphics[width=0.98\columnwidth]{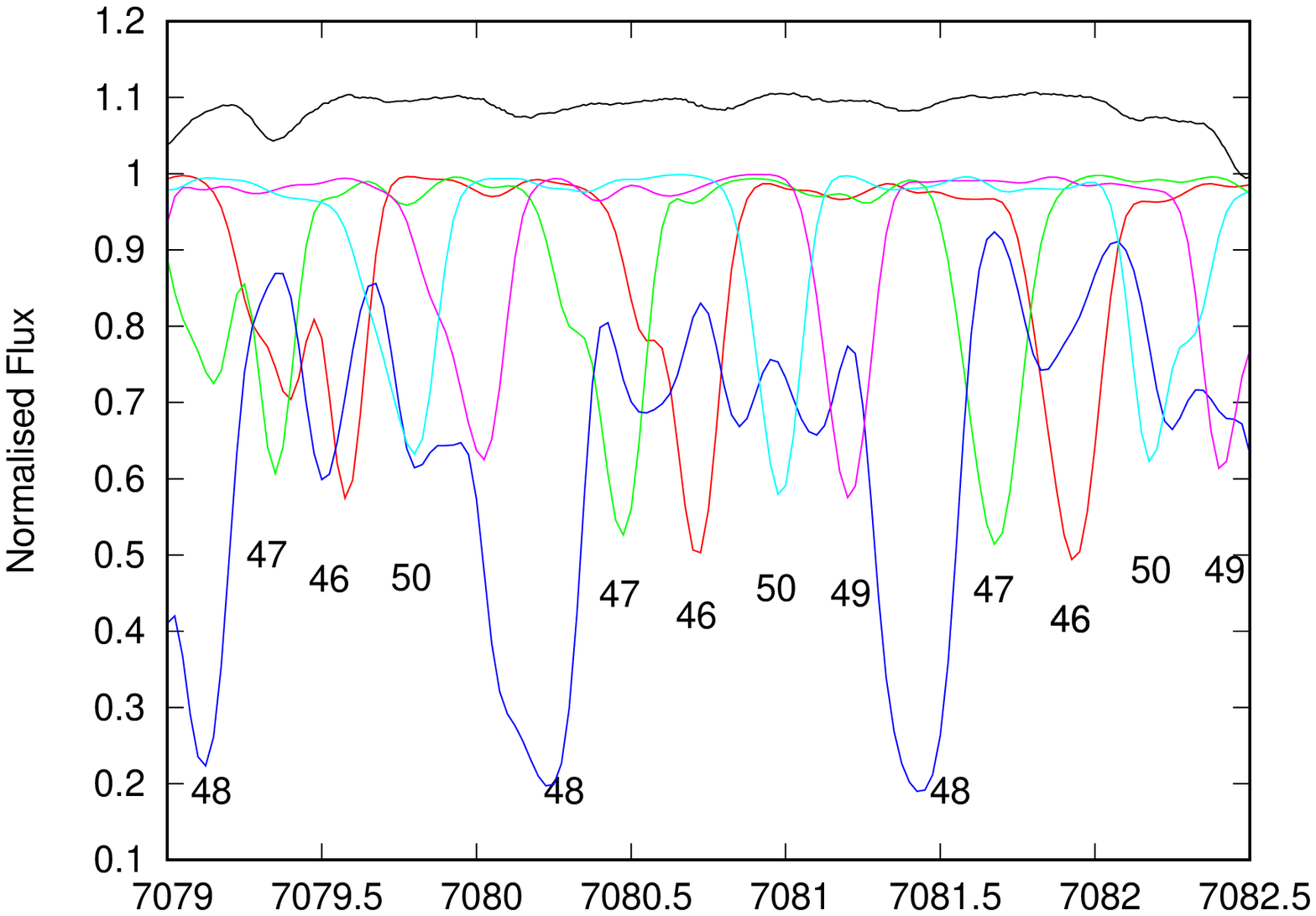}
    \includegraphics[width=0.98\columnwidth]{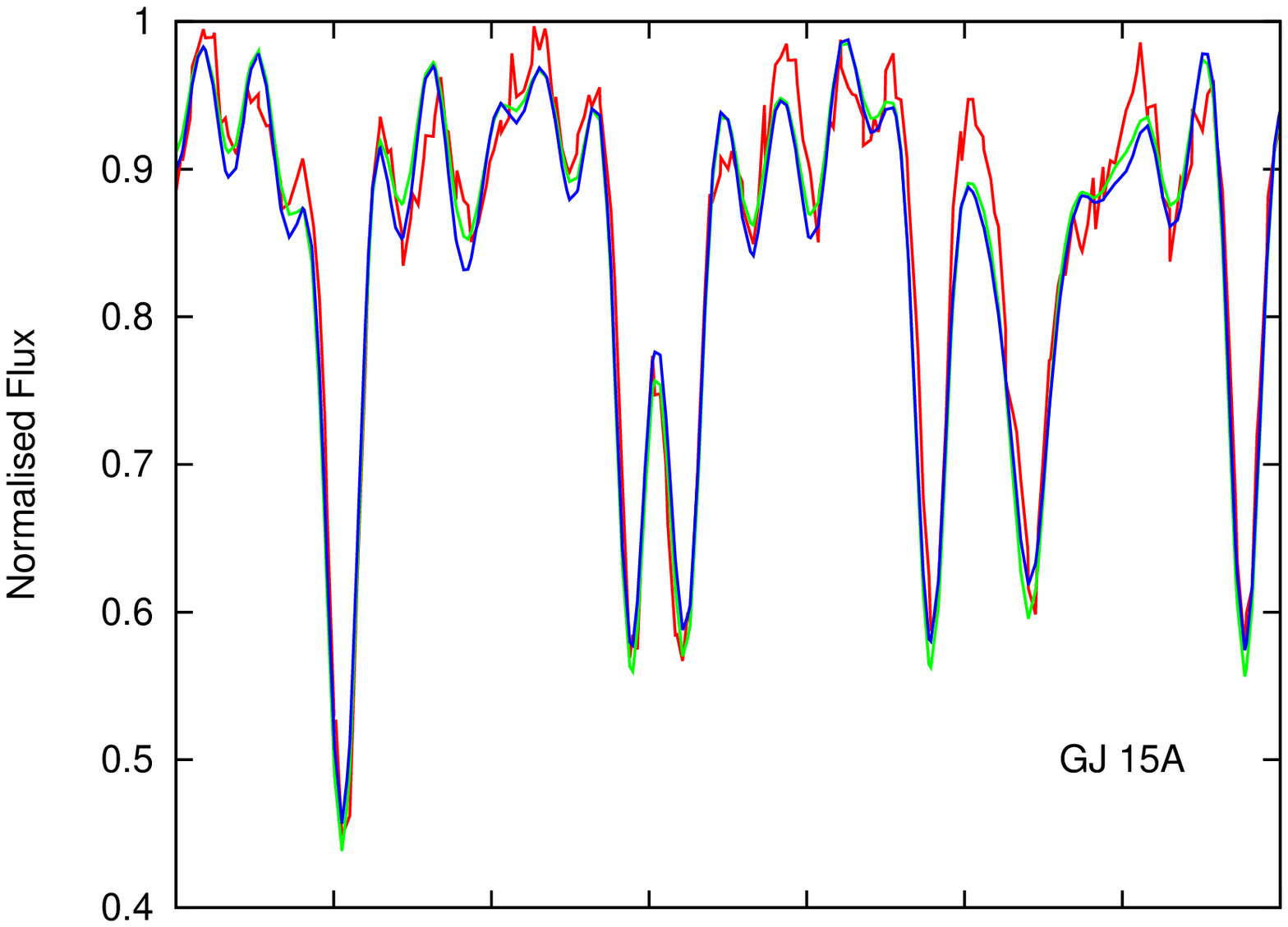}
    \includegraphics[width=0.98\columnwidth]{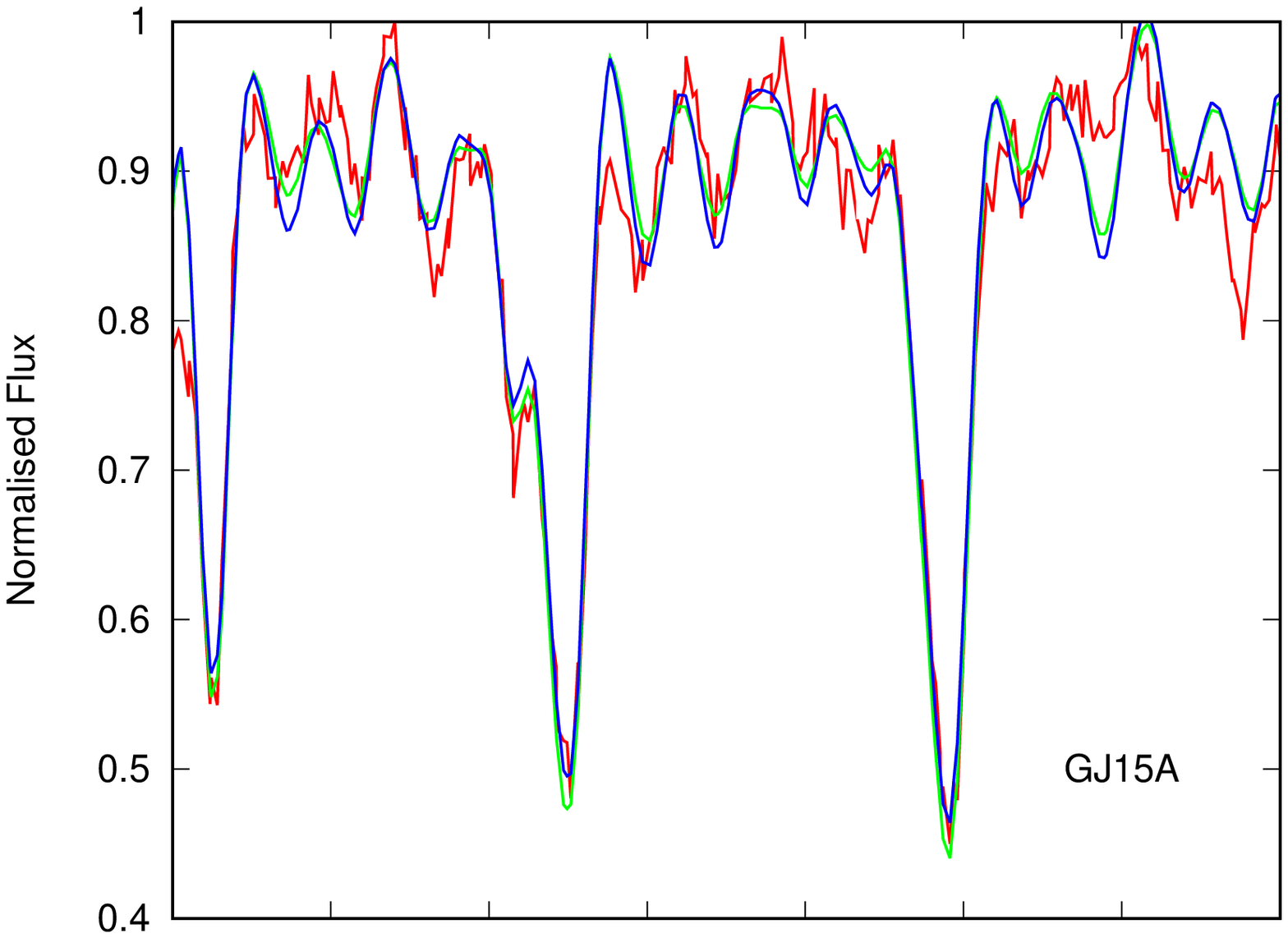}
    \includegraphics[width=0.98\columnwidth]{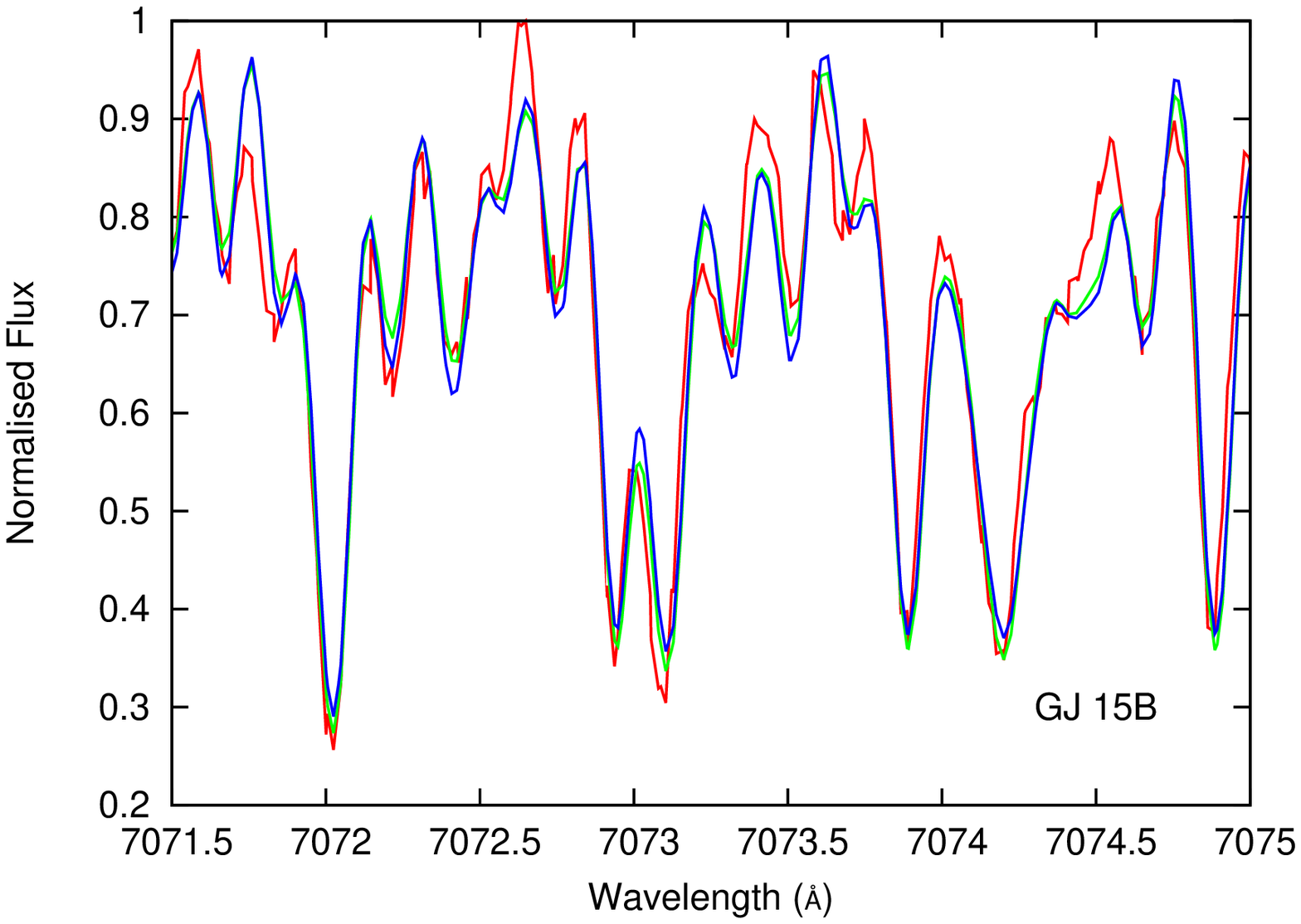}\hskip 0.75 cm
    \includegraphics[width=0.98\columnwidth]{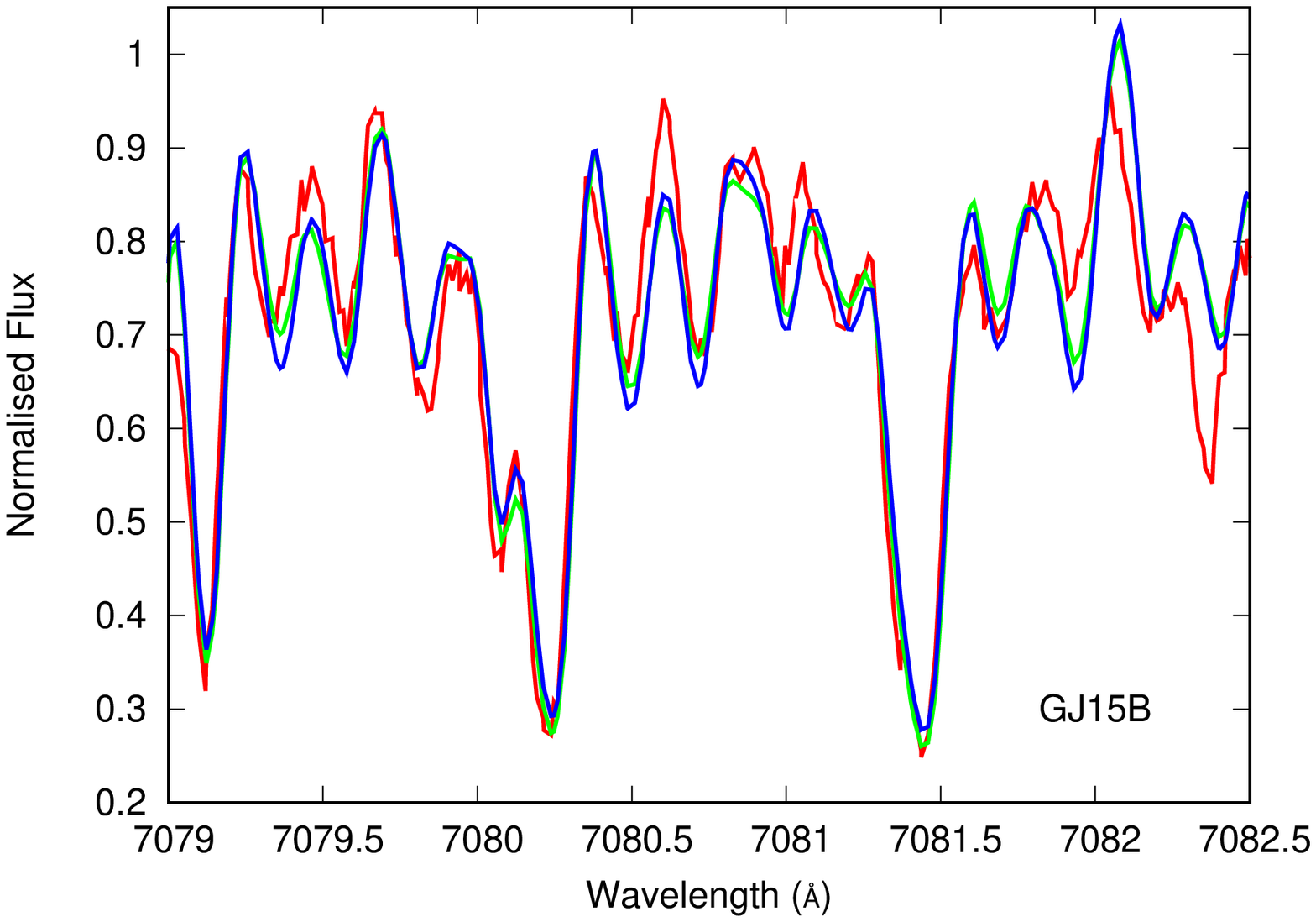}
\caption{\label{_x1} {\it Upper panels:} Features created by the TiO isotopologues in the
spectral ranges used by \cite{lamb77, cleg79} to determine the Ti isotopic abundance in M-stars.
Theoretical spectra of TiO isotopologues were computed for 3500/5.0/0.0 model atmosphere
and solar Ti isotopic ratios. Synthetic spectra are convoluted with R=70,000. The strongest features are marked by
atomic numbers of the Ti isotope in the associated TiO molecule.
Blak line shows the Arcturus spectrum  \citep{hink95}.
{\it Middle panels:} Fits of theoretical spectra computed for 3800/5.0 model atmospheres to the observed spectra of GJ 15A. {\it Bottom panels:} Fits of theoretical spectra computed for 3500/5.0 model atmospheres to the observed spectra of  GJ 15B.
Observed spectra are denoted by red lines, spectra computed for the solar isotopic abundance are shown
by green lines, and spectra computed for M25 isotopic ratios by blue lines.
}
\end{figure*}

As shown in Section \ref{_xxx1}, the heads of the (0,1) band of the $\gamma$-system of TiO isotopologues
form an unresolved blend at 7054 \AA, even in HIRES spectra obtained with R $\sim$ 70,000. Some authors
used  spectral regions beyond 7054 \AA\ for TiO isotope abundance determination.
Spectral ranges $\sim$ 7071.5 -- 7075 \AA\ and 7079 -- 7082.5 were used by \cite{lamb77} and \cite{cleg79} to analyse
spectra of M-stars. Detailed identification of positions of TiO isotopologues is given in their papers.
However, we see from their identifications that at nearly all wavelength there is a contribution from several
TiO isotopologues. To illustrate this point, the two upper panels of Fig. \ref{_x1} show spectra
of all TiO isotopologues computed separately. In fact, we show here fluxes which were used to create the
upper left panel of Fig. \ref{_2945}; at larger scale, these fluxes are shown for the above spectral
ranges in the wavelength frame reduced to $V_r$ = 0.
The strongest features are marked by atomic numbers $z$ of Ti atoms formed by corresponding TiO molecules.

The lower panels of Fig. \ref{_x1} shows fits to spectra of GJ 15A and B for the cases of the ``solar''
isotopic ratios of Ti, i.e. \tsi:\tse:\tei:\tni:\tfi= 8.3 : 7.4 :73.7 : 5.4 : 5.2 \citep{biev93, lodd09},
and a ``modified'' Ti isotopic ratio: \tsi:\tse:\tei:\tni:\tfi= 10.4 : 9.3: 67.0: 6.8: 6.5, in the latter case
abundances of \tei\ were reduced by 10\%, abundances of the other Ti isotopes were increased by 25\%.
For simplicity we label this modified TiO isotopic abundances as the M25 ratios.
The comparison of two spectra illustrates
the sensitivity of the selected spectral features on changes of Ti isotopic ratio.

We note that all isotopologues features can be clearly identified in the observed spectra.
However,  we see only a rather marginal response in these features to the change in Ti isotopic ratios. In all cases
we see blends of a few lines of different isotopes, and the dominant contribution of {\tei}O.
The lines of
{\tei}O are saturated which means they show a rather weak response to isotopic abundance changes.
A qualitative, by-eye estimates of the comparison of the fine details between the observed and computed spectra
seems to suggest a better fit for the case of  ``solar'' Ti isotope ratios.
These fits  adopted [Ti] =  $A_{\odot}(Ti)-A_*(Ti)$ = $-0.1$, where $A_{\odot}(Ti)$, $A_*(Ti)$ are
abundances of Ti in the atmospheres of the Sun and star, for GJ 15A and $-0.2$ for GJ 15B.

 Fits to the  $x_1$ spectral range for spectra of GJ 15A and GJ 15B using three line lists for TiO isotopologues are presented:
S98 --\cite{schw98}, P12 -- updated Plez (1998) and  \LLname{} of the Exomol group \citep{jt760}.
Plots with the fits
are given in Figs. \ref{_cAB} of Appendix which show fits to the observed spectra across  the selected regions also illustrated in Figs.
\ref{_x1}. We performed fits to the head of $\gamma$ system at 7054 \AA. Computed $\chi^2$ minimisation sums are given in table \ref{_clist}.
In all cases \LLname{} provides the best fit.

The following conclusions were also made from the comparisons. \\

a) Positions of the strong absorption feature formed by absorption of photons across \tei\ transitions agree well, but the intensities disagree in some cases.

b) We see more pronounced differences in the positions and intensities for lines belonging to other, less abundant TiO isotopologues.

c) Formally computed $S$ parameters obtained using of different line lists differ from each other, see Table \ref{_clist}.

 We note that this, rather narrow and phenomenological,  comparison of results obtained with different line lists
cannot provide a final conclusion about which is the best one.
However,  our conclusions are valid for  three spectral ranges, which have been used by different authors for Ti isotopic abundance analysis.

However, if we combine
our findings with the more sophisticated, deeper and independent analysis made by \cite{jt760} and \cite{bern20}, we find general agreement that the \LLname{} is the best one.
Indeed, even by-eye inspection confirms that the \LLname{} line list is better at reproducing the observed features in terms of both line intensities and line positions,
see Fig. \ref{_clist}.

\begin{table*}
\caption{\label{_clist} Parameter $S$ computed across fitting ranges given as wavelengths in \AA.}
\begin{tabular}{lcclll}
\hline
\hline
\noalign{\smallskip}
    & 7053 -- 7060          &    7071 -- 7084   &    7580 -- 7594     \\
\noalign{\smallskip}
\hline
\noalign{\smallskip}

P12 &  0.73 $\pm$ 0.03 (A)  & 2.75$\pm$ 0.04    &   0.75 $\pm$ 0.02    \\
    &  2.11 $\pm$ 0.05 (B)  & 7.60$\pm$ 0.08    &   3.50 $\pm$0.04     \\
S98 &  0.66 $\pm$ 0.03 (A)  & 2.45$\pm$ 0.04    &   0.76 $\pm$ 0.02    \\
    &  1.64 $\pm$ 0.04 (B)  & 3.64$\pm$ 0.05    &   3.40 $\pm$0.04     \\
\LLname{} &  0.47 $\pm$ 0.02 (A)  & 1.10$\pm$ 0.03    &   0.73 $\pm$ 0.02    \\
    &  1.28 $\pm$ 0.04 (B)  &  2.32$\pm$ 0.04   &   3.19 $\pm$0.04     \\
\noalign{\smallskip}
\hline
\end{tabular}
\end{table*}

\subsubsection{Spectral range $x_2$}
\label{_xx2}

The spectral range $x_2$ is favorably different from $x_1$. Heads of the TiO bands  are well spaced across
the wavelengths which improves their identification and analysis. This is more relevant for the
band heads of {\tfi}O and \tni O which are located before the head of the strong {\tei}O band.
Despite the strong blending between the {\tfi}O band head and an Fe I line, the spectral detail is clearly seen in
the observed spectrum, which is relevant to stars with low \vsini.

Unfortunately, in spectra of the two stars the \tsi O band head coincides with strong telluric feature.
Still, molecular band heads of other TiO isotopologues are seen, therefore we were able to determine all ratios,
\tsi~ was determined as 100 - (\tse + \tei + \tni + \tfi). Moreover,
we include [Ti] in the minimisation process due to the dependence of isotopic ratios  on the abundance of element,
see fig. 2 in \cite{pavl20} for the case of the carbon isotopic ratio determination. Indeed, strong and
weak features show different responses to abundance variations.

We performed  isotopologues abundance analysis using the determination of min $S$(\tse, \tei, \tni, \tfi, and [Ti])
for the set of synthetic spectra computed for the grid of input parameters:

-- \tse\ varies in the range [3.0 -- 10.0] with step 1.0;

-- \tei\ varies in the range [71.5 -- 77.5] with step 0.5 ;

-- \tni\ varies ithe range [3.0 -- 10.0] with step 1.0;

-- \tfi\ varies in the range [3.0 -- 10.0] with a step 1.0;

-- [Ti] varies in the range [0 -- 0.1] and [0.1 -- 0.2] with a step 0.02 for  GJ 15A and GJ 15B, respectively.

These parameter ranges were determined in a set of numerical experiments.

Our 5D minimisaton procedure provides a set of minimisaton sums, $S$, followed by formally computed errors, $\Delta S$.
We obtained $S_{min}$ = 1.030 $\pm$ 0.021 for GJ 15A, and then averaged solutions falling
in the window 1.030 $< S <$ 1.052. For  GJ 15B we average solutions in the range
3.332 $ < S < $3.375. Due to higher level of noise in GJ 15B spectrum our solution
is less confident for this cooler star. Table \ref{_xAB} shows relative abundances of TiO isotopologues determined from the $\chi^2$ fits to
the observed spectra of GJ 15A and GJ 15B.

\begin{figure*}
   \centering
    \includegraphics[width=0.98\columnwidth]{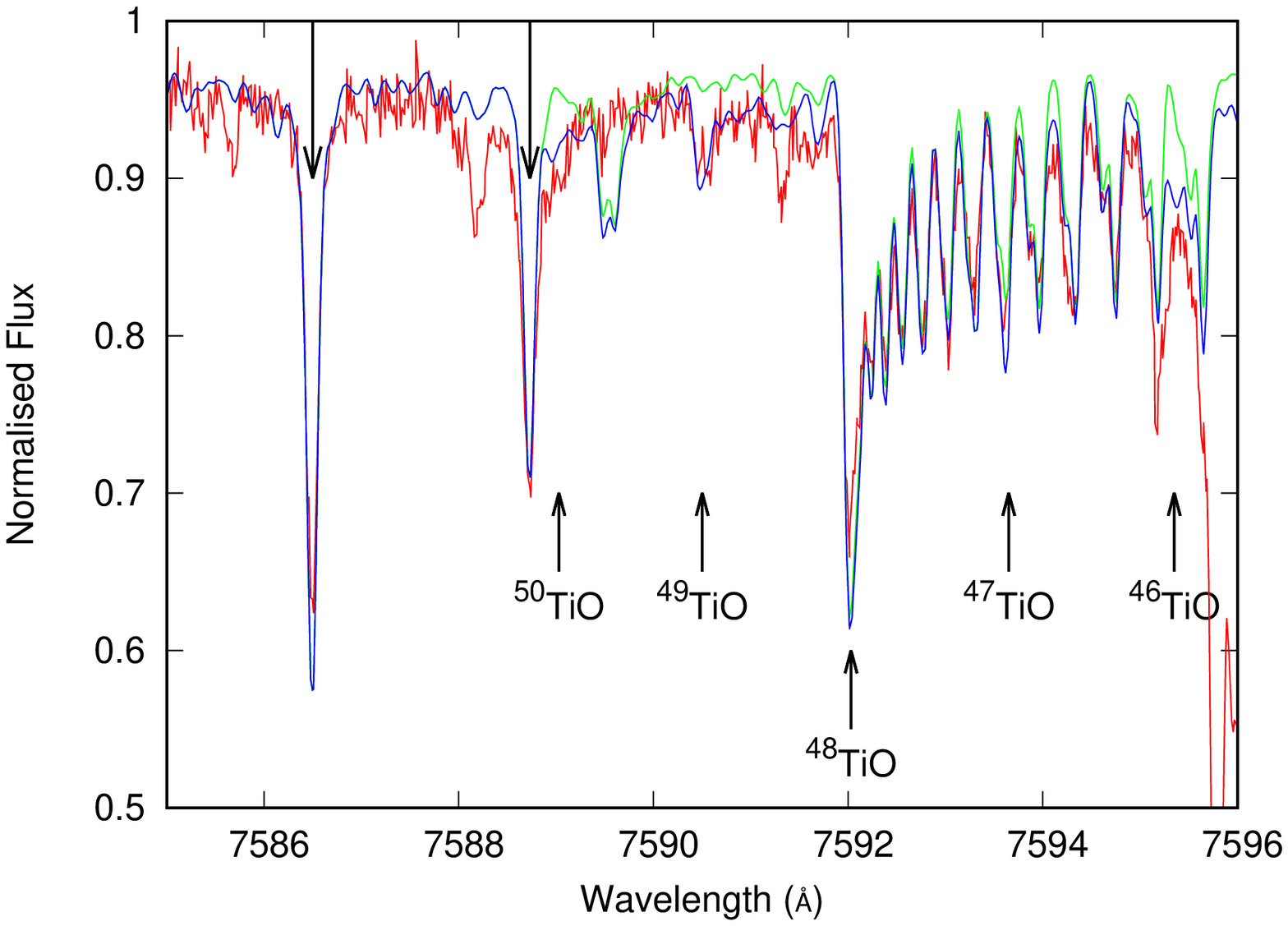}
    \includegraphics[width=0.98\columnwidth]{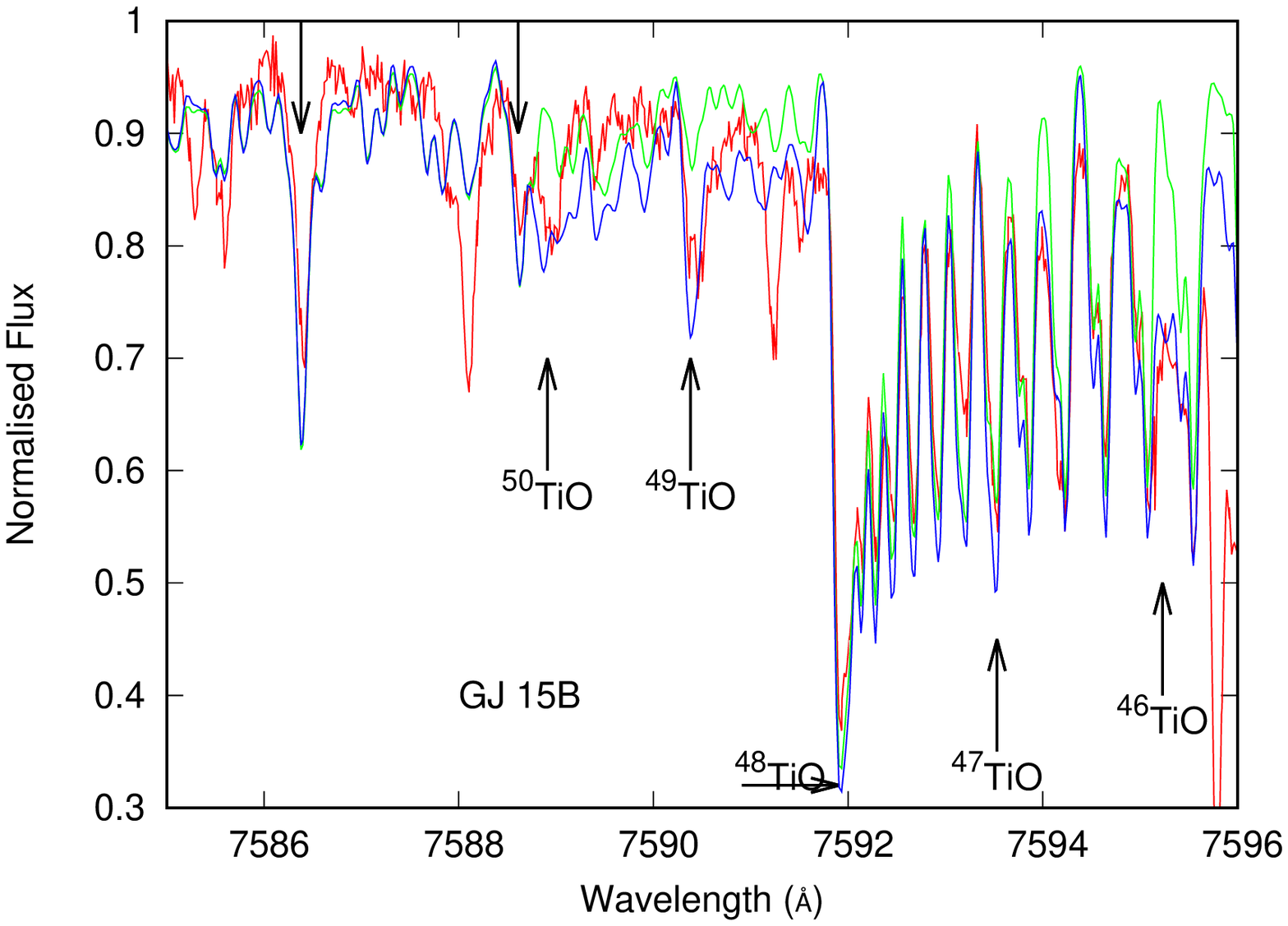}
\caption{\label{_xf}  Fits of our synthetic spectra to the observed
 GJ 15A {\it (left panel)} and GJ 15B {\it (right panel)} spectra with Ti isotopic
 ratios \tfi: \tni: \tei: \tse: \tsi =
 6.2:7.9:72.8:5.2:7.9 and 6.0:5.8:76.6:4.2:7.4, respectively, shown by blue lines.
Observed spectra are shown by red line, fluxes computed with
 account of only \tei O shown by green line.  Downward arrows mark Fe I lines.} 
\end{figure*}

The green lines in Fig. \ref{_xf} show spectra computed without
\tsi O, \tse O, \tni O and \tfi O.
Comparison with the ''full`` sample of TiO isotopologues shows the contribution of the less abundant species to the total spectrum.

Generally speaking, in the ideal case of isolated stars of
similar masses formed a wide binary system we do not expect large
differences in their abundances.
As a result any differences in isotopic ratios may be caused by other reasons, which might tell us
something about our procedure  or the physical state of the stars. Here,
we find good agreement in the fits to the \tni O and \tse O features, but we also see
some disagreement for \tfi O in the
shape and intensity between the computed molecular bands and the  observations.

\begin{table}
\centering
\caption{\label{_xAB} Isotopologues abundances and Ti abundance in atmospheres of GJ 15A and GJ 15B. }
\begin{tabular}{cccc}
\hline
\hline
\noalign{\smallskip}
Isotopologue & the Sun & GJ 15A &  GJ 15B \\
\noalign{\smallskip}
\hline
\noalign{\smallskip}

\tsi &  8.3  & 7.9  $\pm$ 0.01   &  7.4  $\pm$ 0.3           \\
\tse &  7.4  & 5.2  $\pm$ 0.01   &  4.2 $\pm$ 0.4    \\
\tei & 73.7  & 72.8 $\pm$ 0.01   & 76.6 $\pm$ 0.2    \\
\tni &  5.4  & 7.9 $\pm$ 0.01   &  5.8 $\pm$ 0.3    \\
\tfi &  5.2  &  6.2 $\pm$ 0.01   & 6.0 $\pm$ 0.0    \\
$[$Ti$]$ &  0.0  & 0.040 $\pm$ 0.001   & 0.199 $\pm$ 0.003 \\
\noalign{\smallskip}
\hline
\end{tabular} \\
\end{table}

To compare the fits of synthetic spectra in the $x_2$ spectral using the different line lists S98, P12 and \LLname{} lists,
$S$ parameter were computed. In these computations we adopted solar isotopic ratios for
the Ti isotopes. The fits to the observed spectra are shown in Fig. \ref{_xxc}; the computed values of  $S$ for all of them are given in Table \ref{_clist}.

We note, that the two Fe I lines seen in this spectral range can be fitted with [Fe] = $-0.4$ for both stars, which confirms
deficit of iron found previously,
see Table \ref{_gj15info}. Our minimisaton procedure
allows us to determine [Ti] = 0.040 $\pm$ 0.001 and 0.199 $\pm$ 0.003 for GJ 15A and GJ 15B, respectively.

\subsubsection{Spectral ranges $x_3 - x_5$}

Unfortunately TiO is weak in the $x_3$ -- $x_5$ spectral ranges for the observed and computed spectra of GJ 15A and GJ 15B.
The  $x_3$ spectral range is dominated  by an Na I line; here we see only a weak  {\tei}O band head.
This spectral range may be more useful for giant spectra, where the  Na I
8200 \AA\ subordinate triplet is much weaker.
Indeed, isotopologue bands are here in ``ascending order'', i.e. from \tsi O to {\tfi}O, which is the opposite order to
the $x_2$ spectral range.

As we noted above, molecular bands of TiO isotopologues in $x_4$ spectral range form a single feature in the observed
and computed spectra, as in $x_1$ spectral range. We cannot distinguish
contributions from TiO isotopologues in the spectra of GJ 15A and GJ 15B as TiO appears much weaker here and there is strong pollution by telluric lines.

In the $x_5$ spectral range we observe a descending order of the isotopologue bands, similar to  $x_2$. However, the TiO
absorption is too weak here to form any notable features which can be used in the analysis.

\section{Discussion}

 Though molecular spectroscopy of TiO provides the most accurate method of determining titanium isotope abundance in cool stars, analysis of TiO spectroscopy is very challenging. As a heavy molecular system, the TiO spectra is complex with spectral features formed by blends of many absoption lines often from different isotopologues. To obtain high accuracy measurements, it is vital to select the best spectral region to do high-resolution analysis.

The main aims of this paper are a) to test the new \LLname{} TiO line list for the determination
of the TiO isotopologue abundances in the observed spectra of cool M-dwarfs, b) to determine the most appropriate
spectral ranges for the determination of Ti isotope abundances in their atmospheres, and
c) to retrieve the Ti isotopic ratios for GJ 15A and GJ 15B to test the performance of this method.

\subsection*{a) Line list accuracy}

Previous comparisons of TiO line lists (e.g. by \cite{jt760}, \cite{bern20}, \cite{piette20}) have concluded that the  \LLname{} line list is generally superior to older line lists. However, these comparisons have not explicitly considered the quality of predicted isotopologue spectra, i.e. how accurately does each line list reproduce the astronomically observed isotopologue frequency shifts and therefore allow the abundance of different titanium isotopes  to be accurately measured astronomically. In this paper, we have modelled several spectral regions that can be used for Ti isotope abundance analysis in cool stars, comparing our model isotopologue spectra computed using  \LLname{} line list with other available line lists. Detailed comparisons with  high resolution M-dwarf spectra  with different line lists
performed in this paper shows that \LLname{} line list performs much better than
the earlier line lists, particularly in matching line positions, 
see Fig. \ref{_cAB}.

\subsection*{b) Best spectral ranges for isotope determination}

In the past, majority of researchers  concentrated on
analysing of the fine details of absorption spectra  formed in the redward degrading (0,0)
band of the TiO $\gamma$ band system. However, observable features here are formed by blends of
many lines belonging to different isotopologues. In our case synthetic spectra computed for the ``solar'' Ti isotopic ratios
 across our $x_1$ spectral range can be fitted to the observed spectra of GJ 15A and GJ 15B; however,  we found here rather weak dependence on
the reasonable changes (25 \%) of TiO  isotopologues abundances.

We investigated a few possible spectral ranges and found that the $x_2$ spectral range is the most useful because the region contains
band heads for four of the five stable TiO isotopologues, enough for the determination of all titanium isotope abundances. Note
that the \tsi O band head being severely blended by telluric feature, but \tsi\ can be determined as \tsi = $100-$(\tse+\tei+\tni+\tfi).

 Our analysis of band heads across the $x_2$ spectral range provides more reliable abundances of TiO isotopologues,
especially of {\tfi}O and {\tni}O, which are not blended by strong lines of {\tei}O. Conversely, this
spectral range is very interesting from the viewpoint that some nucleosynthesis computations predict large
deviation of the \tfi\ and \tni\ from the ``solar'' abundances  at the early epochs of our Universe's evolution,
see Introduction. 

 We developed a new procedure allowing to determine all Ti isotope abundances together with Ti abundance using this spectral region.
To find the best solution we applied a  5D minimisation procedure with input parameters \tse, \tei, \tni, \tfi\ and
titanium abundance [Ti] to take into account the dependence of determined isotopic ratios on the the adopted titanium abundance.
In that way we determine the titanium isotopic ratios and abundance of Ti from our analysis of $x_2$ spectral range
in the framework of the self-consistent approach. This a the key part of our procedure developed for  TiO
isotopologue abundances analysis. Furthermore, to get statistically significant results we implemented a special procedure as part of our solution determination. Namely, the set of the best solutions were sorted in the ascending order of the minimisation sums $S$, correspondent isotopic ratios and Ti abundances were averaged across
$S_{\rm min}-\Delta S \leq S \leq S_{\rm min}$, where $S_{\rm min}$ and $\Delta S$ are the minimal $S$ and the formal error of its determination $S$,
see section \ref{_fos}.

\subsection*{c) Analysis of GJ 15A and GJ 15B spectra}

 In our study, we model TiO spectra of two stars which form a binary system, specifically the CARMENES spectra of two M-dwarfs, i.e GJ 15A and GJ 15B, in the spectral ranges of our interest. We selected GJ 15A and GJ 15B because a) they are comparatively well studied and b) they are slow rotators.
We anticipated that the isotope abundance in both stars should be equal and thus we could use the binarity of the system to prove the reliability of our results in the case that a) we know all abundances in atmospheres of both stars, b) their effective temperatures are well determined, to say, from the fits to observed spectral energy distributions; c) model atmospheres are computed for the found \Tef, \logg and abundances, d) spectra of both components were observed with the same high quality. Unfortunately, none of this conditions are entirely satisfied. We fixed \Tef =3800 K for GJ 15A, and \Tef =3500 K for GJ 15B, and adopt \logg = 5.0 for both stars, because the system is rather old, see Table 1. Furthermore, GJ 15B is fainter, obtained spectrum for the cooler dwarfs is of lower S/N, in comparison with GJ 15A.

Our best fits to the observed spectra of GJ 15A and GJ 15B across the $x_2$ spectral range provide
non-solar Ti isotopic ratios of  \tsi:\tse:\tei:\tni:\tfi\  = 7.9 : 5.2 : 72.8 : 7.9 : 6.2  for GJ 15A (M1 V) and
7.4 : 4.2 : 76.6 : 5.8 : 6.0 for GJ 15B (M3 V) with accuracy of order $\pm$  0.2. Furthemore, [Ti] = 0.040 and 0.199 with accuracy $\pm$ 0.10 were also
determined for GJ 15A and GJ 15B, respectively.
The differences in the isotope abundances between the two stars is significantly larger than the uncertainties of each individual determination.  It is unclear whether these differences reflect actual differences in isotope abundances in the two stellar objects, inaccuracies in the determination of the relevant physical properties of both systems (e.g. temperature) or can be attributed to modelling uncertainties. For more definite conclusions, abundance analysis for all elements should be considered in both stars. Ideally, observed spectra of both M-dwarfs should be of the same quality, i.e. of the same resolution and S/N.

We note some disagreement in the shape and intensity between the computed molecular bands of \tfi O and the observations. Likely, here we have some contribution of other unidentified molecule(s). Given the promise of this spectral region in determining titanium isotope abundances, we suggest detailed consideration of this issue in the future.

We anticipated that pecularities in abundances can be explained by the inaccuracies of $T_{\rm eff}$ determination, e.g.
we found that lowering \Tef\ of GJ 15A by 100~K reduces the Ti abundance for the star by $-0.15$.
Furthermore, the Ti I line seen in lower panels of Fig. \ref{_AAB}
can be fitted by adopting [Ti]=$-0.2$. However, this Ti abundance describes better the Ti line in GJ 15A. To describe Ti line in GJ 15B
spectrum we should adopt larger titanium abundance. Furthermore, from the fit to $x_2$ spectral range we obtained larger [Ti] in the atmosphere of
the GJ 15B.

Of note, we found that the binary seems to be Fe deficient, as previously reported by several authors,
see Table \ref{_gj15info}. Indeed, to get reasonable fits to the two Fe I lines observed in $x_2$ spectral range
we adopt [Fe] =-0.4, see Fig. \ref{_xf}.  Conversely, we obtain rather solar abundance of Ti in atmospheres of two stars. Temperature changes cannot explain the observed overabundance of Ti in respect to Fe. The overabundance of Ti in respect to Fe can be interpreted, at least qualitatively, due to $\alpha$-elements abundance enhancement in metal-poor stars, see Introduction.

\section{Conclusions}
 Accurate determination of titanium isotope abundances can shed new light on astrophysical processes. On the large scale, comparing isotope abundances in atmospheres of the slow evolving M-dwarfs to abundances in more massive, fast evolving M-giants could provide the new information about evolutionary processes in our Galaxy. In individual astrophysical systems, such as the Earth-Moon system and host star-exoplanet systems, more local histories can be determined, e.g. near-identical titanium (and oxygen) isotope abundances on Earth and the Moon challenge the dominant theory of the formation of the Moon after a giant impact, see Introduction.

Our results demonstrate that titanium isotope abundances in M stars can be most effectively measured using high-resolution spectroscopy of TiO in the $x_2$ spectral region,  i.e. air wavelengths of 7580 -- 7594 \AA, through fitting observed spectral intensities of many isotopologue features against models computed using the \LLname{} line list. The successes and limitations of this methodology were explored through detailed studies of the spectroscopy of the GJ 15A and GJ 15B binary system. We find that for optimal results \Tef, \logg, abundances of all elements should be determined by the most accurate procedure and the  spectra of all M-dwarfs should be of the same, very high quality. Taking into account all uncertanties of our analysis we conclude that our determinations is accurate to $\sim$ 5\% absolute abundance
of main isotopologue and 20-30\% relative abundances of minor isotopologues. We believe, that the use of better quality spectra will increase
these accuracies by factor of 10. Fortunately, the spectral range of our interest, i.e. $x_2$, lies in the near-infrared, where observed fluxes
are larger than in optical wavelengths, and detectors provide data of higher S/N. 

\section*{Acknowledgements}

This study was funded as part of the routine financing
programme for institutes of the National Academy of Sciences
of Ukraine.
Spectroscopic data calculated by the
ExoMol group (funded by ERC as part of the Advanced Investigator 267219 project and UK
Science and Technology Research Council (STFC) No. ST/R000476/1), the SIMBAD database (CDS,
Strasbourg, France), and the Gaia spacecraft data (European Space Agency) were used. This study is based in part
on archival data obtained using the infrared telescope operated by the University of Hawaii under a cooperative
agreement with NASA. Authors would like to thank the SAO/NASA ADS team for the development and support
of this remarkable data system.
This work has made use of data from the European Space Agency (ESA) mission
{\it Gaia} (\url{https://www.cosmos.esa.int/gaia}), processed by the {\it Gaia}
Data Processing and Analysis Consortium (DPAC,
\url{https://www.cosmos.esa.int/web/gaia/dpac/consortium}). Funding for the DPAC
has been provided by national institutions, in particular the institutions
participating in the {\it Gaia} Multilateral Agreement. We thank CARMENES team for providing
high quality database of M-dwarf spectra which was used in this paper.
We thank the anonymous referee for a thorough review and we highly
appreciate the comments and suggestions, which significantly contributed
 to improving the quality of the paper.

\bibliographystyle{aa}
\bibliography{mnemonic,1,yp,journals_astro,jtj,partition}
\newpage

\appendix
\section{}
\begin{table}
\caption{An example of input data of \tse O lines in our computation}
\label{input}
{\tt
\begin{tabular}{ccc}
\hline\hline
\noalign{\smallskip}
$\lambda$ (A) in air & $gf$  & $E''$ (eV)\\
\noalign{\smallskip}
\hline
\noalign{\smallskip}
 3493.707& 3.172E-05 & 0.890 \\
 3493.707& 3.172E-05 & 0.890 \\
 3494.907& 3.323E-05 & 0.897 \\
 3494.907& 3.323E-05 & 0.897 \\
 3496.134& 3.481E-05 & 0.903 \\
 3496.134& 3.481E-05 & 0.903 \\
 3496.479& 3.521E-05 & 0.917 \\
 3496.479& 3.521E-05 & 0.917 \\
 3497.389& 3.647E-05 & 0.910 \\
 3497.389& 3.647E-05 & 0.910 \\
... \\
\hline\hline
\end{tabular}
}
\end{table}

\begin{figure*}
   \centering
   \includegraphics[width=0.98\columnwidth]{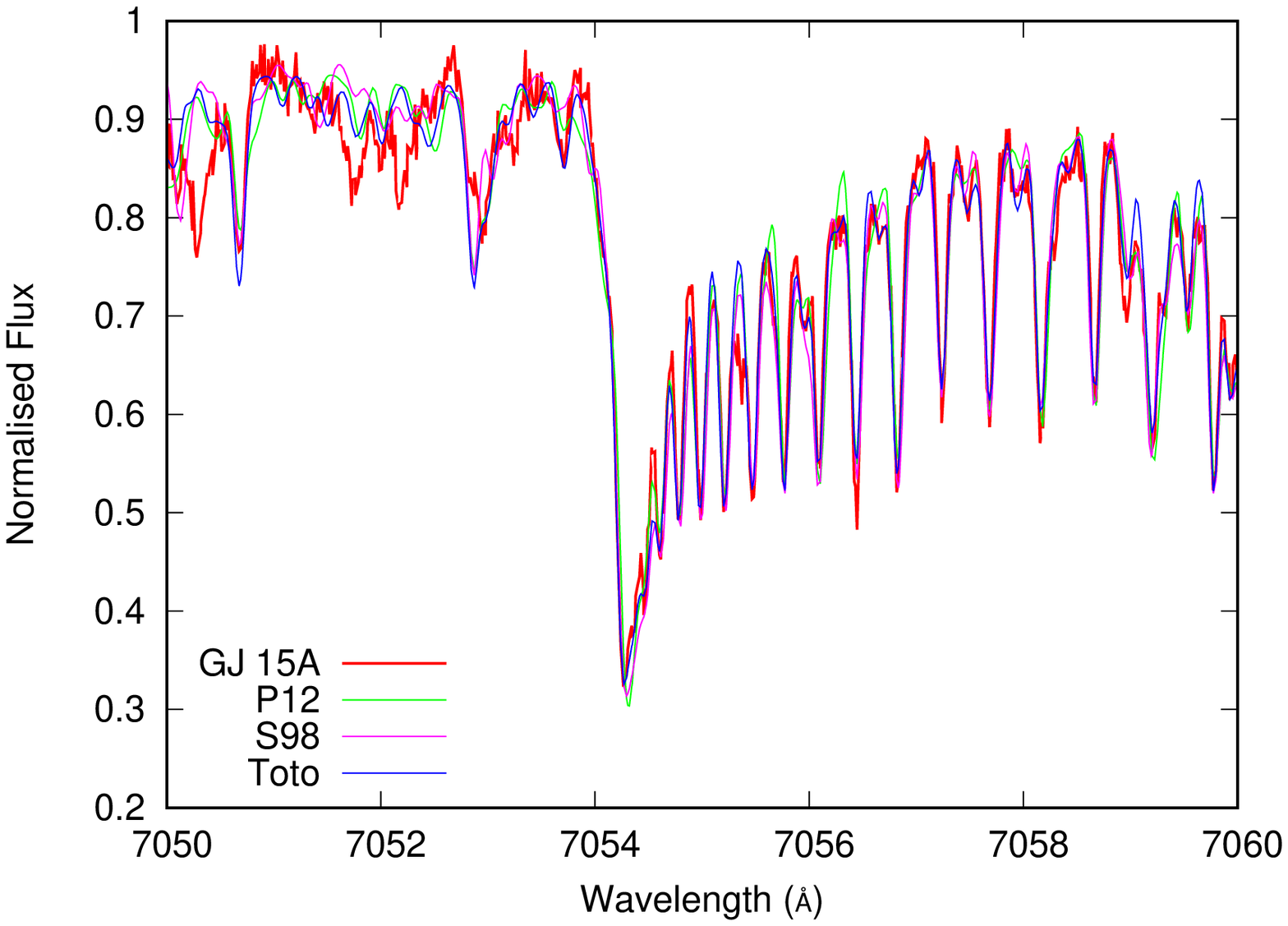}
\includegraphics[width=0.98\columnwidth]{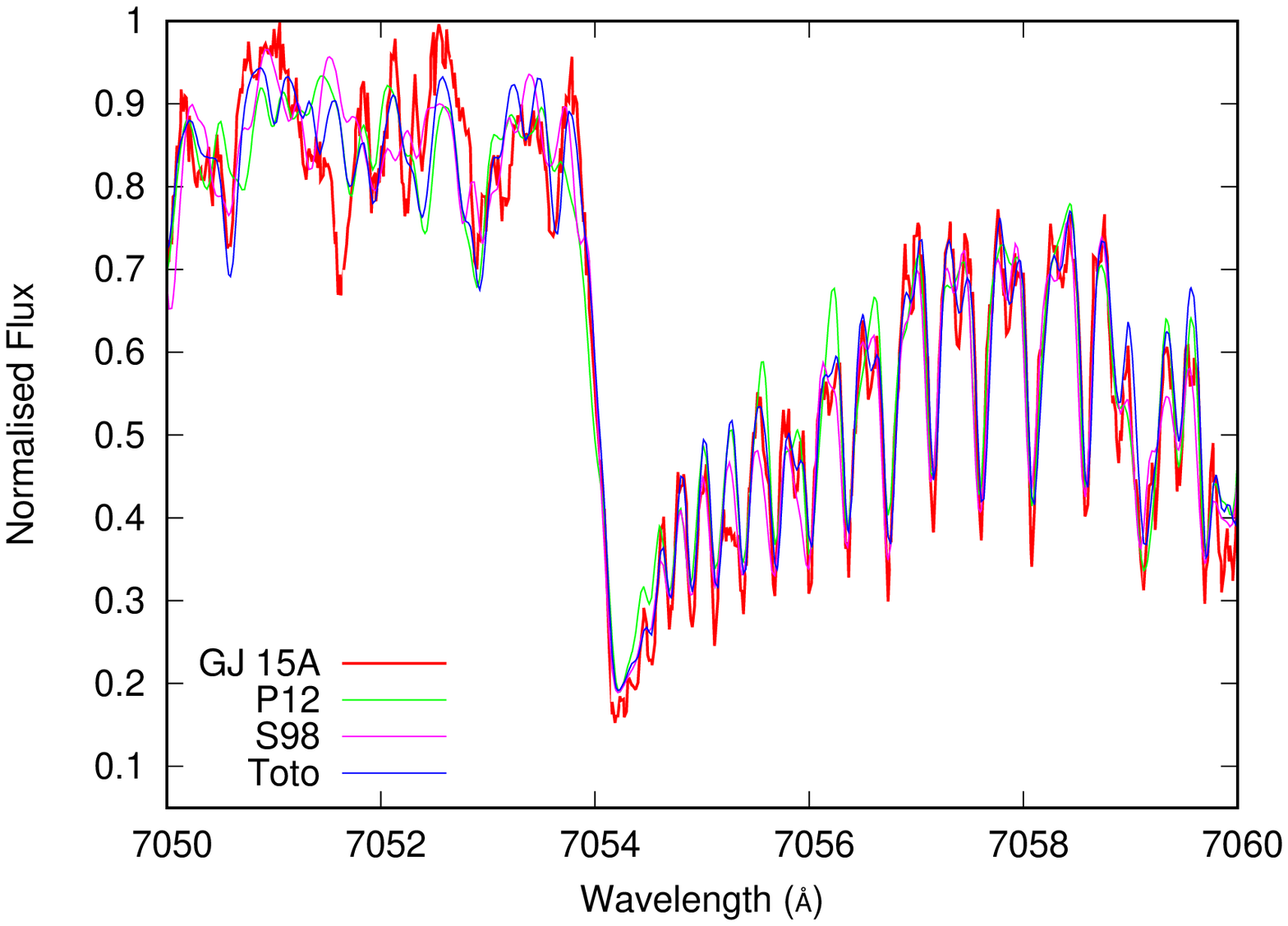}
    \includegraphics[width=0.98\columnwidth]{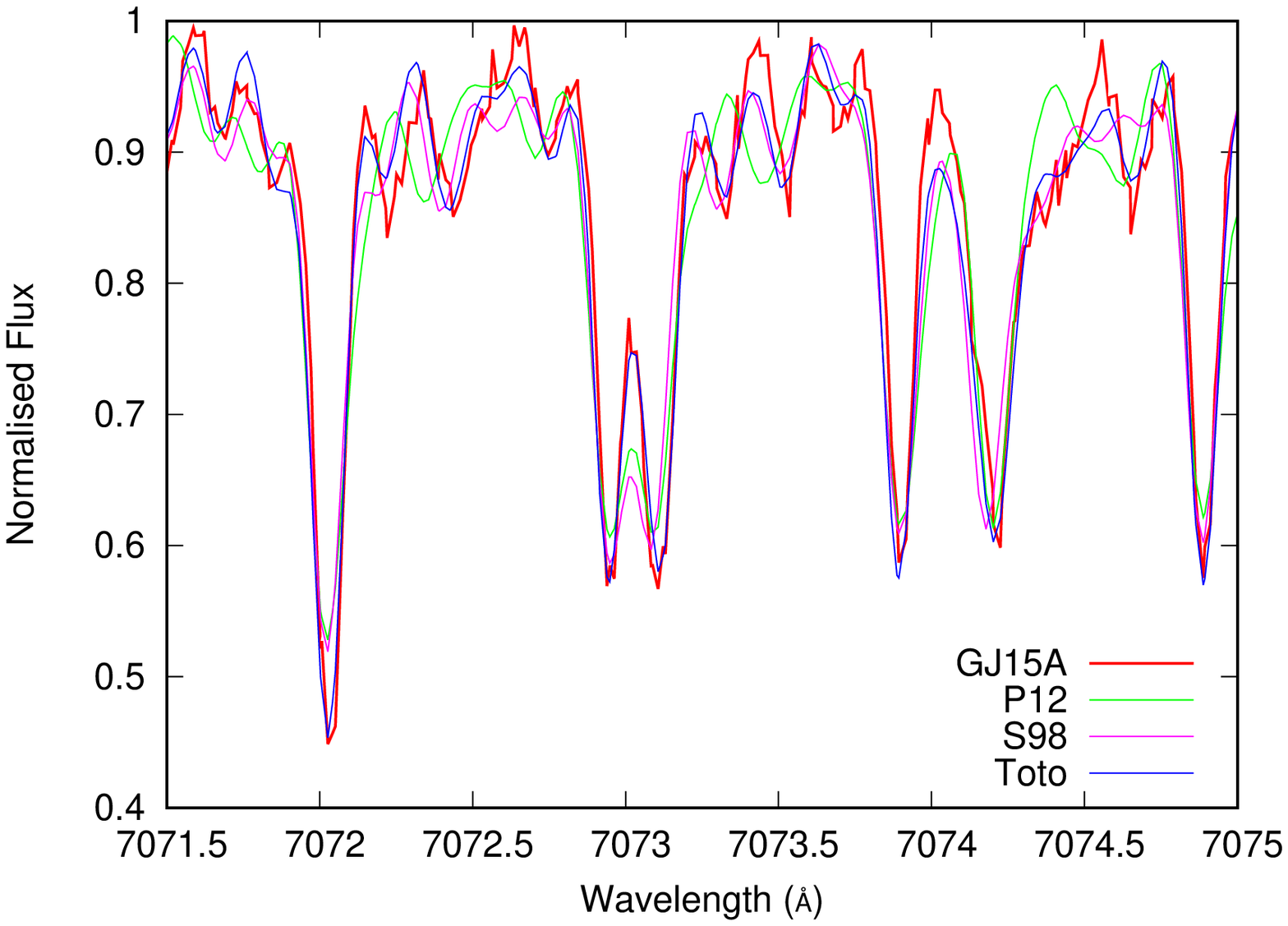}
    \includegraphics[width=0.98\columnwidth]{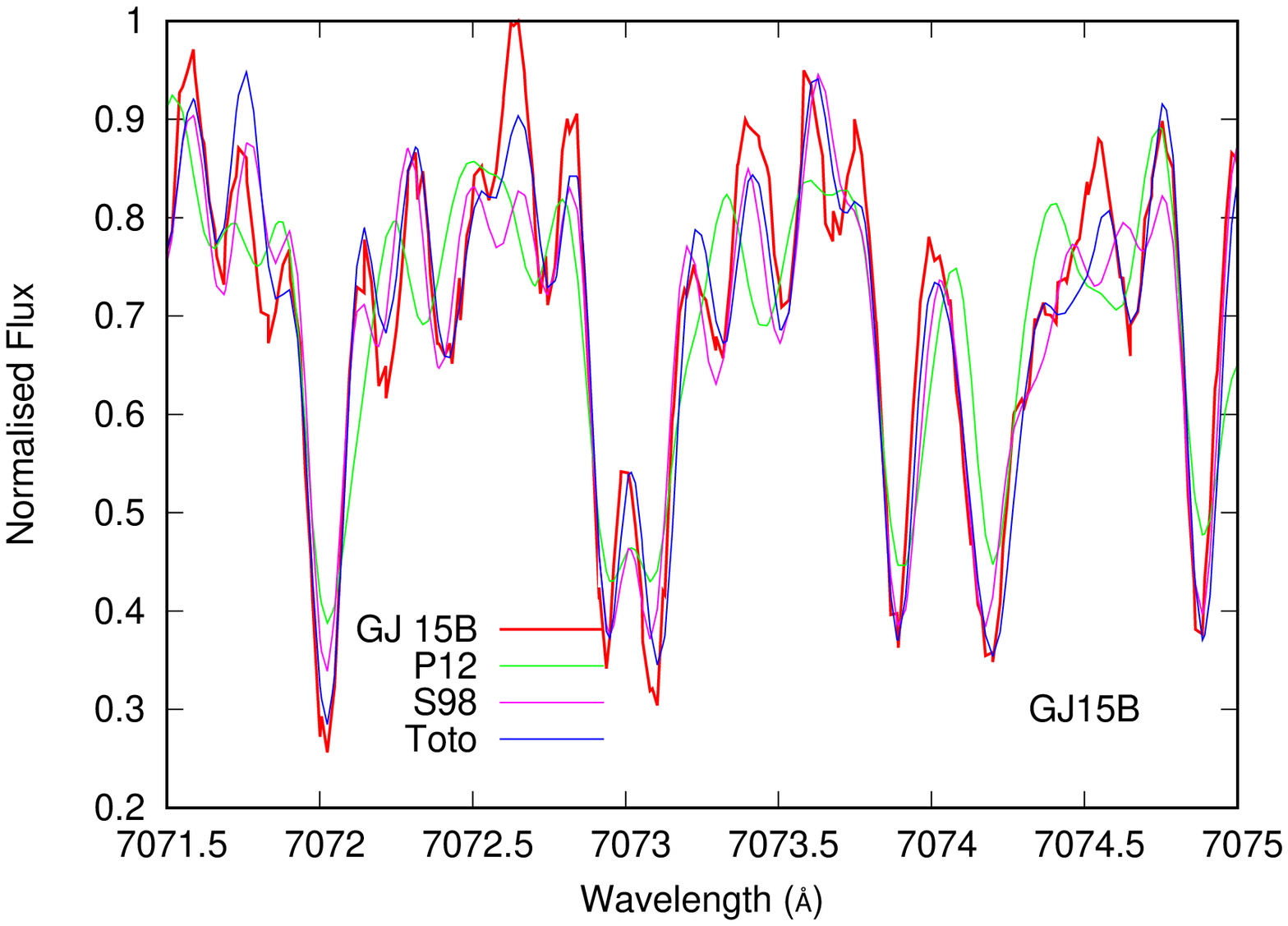}
    \includegraphics[width=0.98\columnwidth]{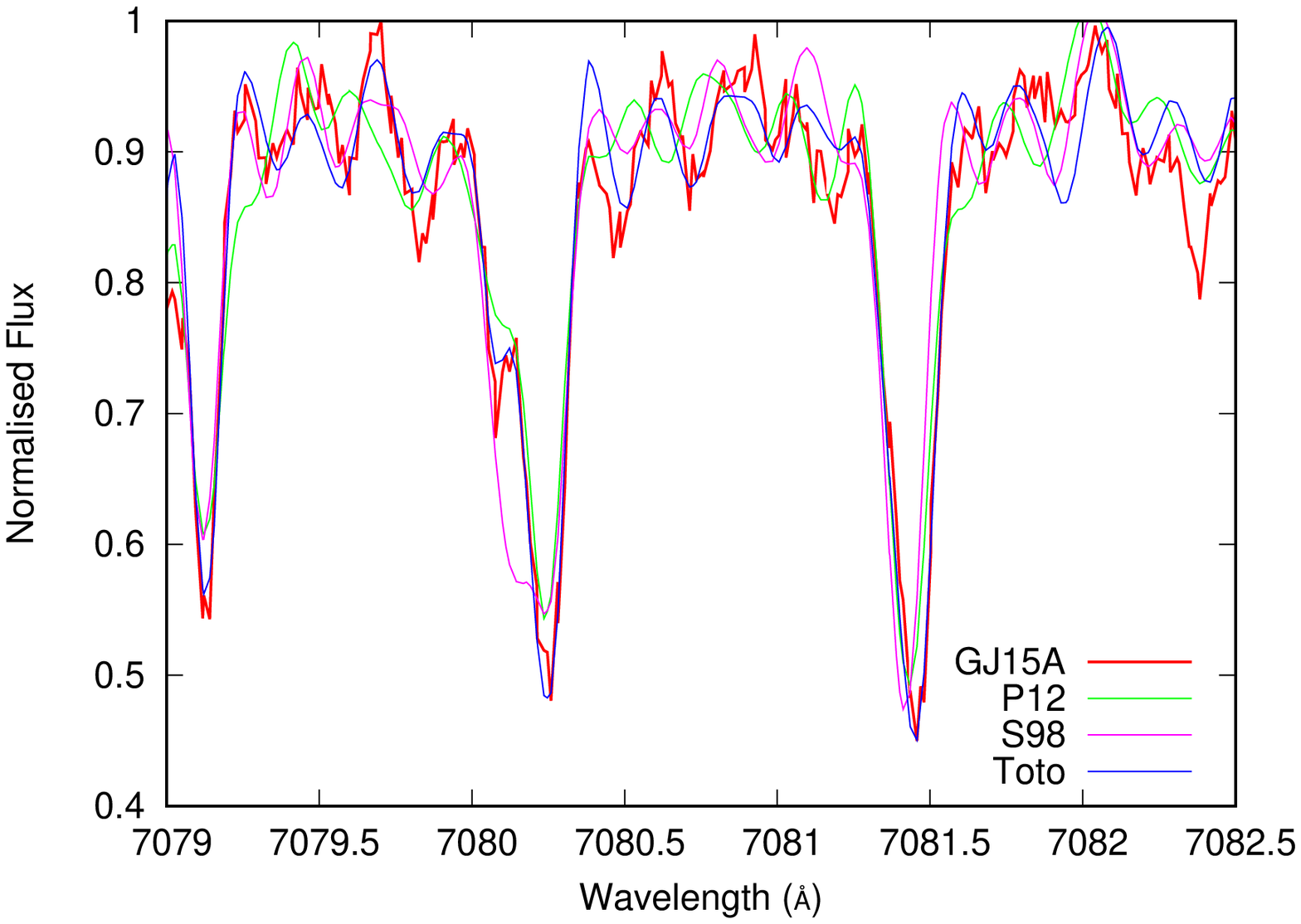}
        \includegraphics[width=0.98\columnwidth]{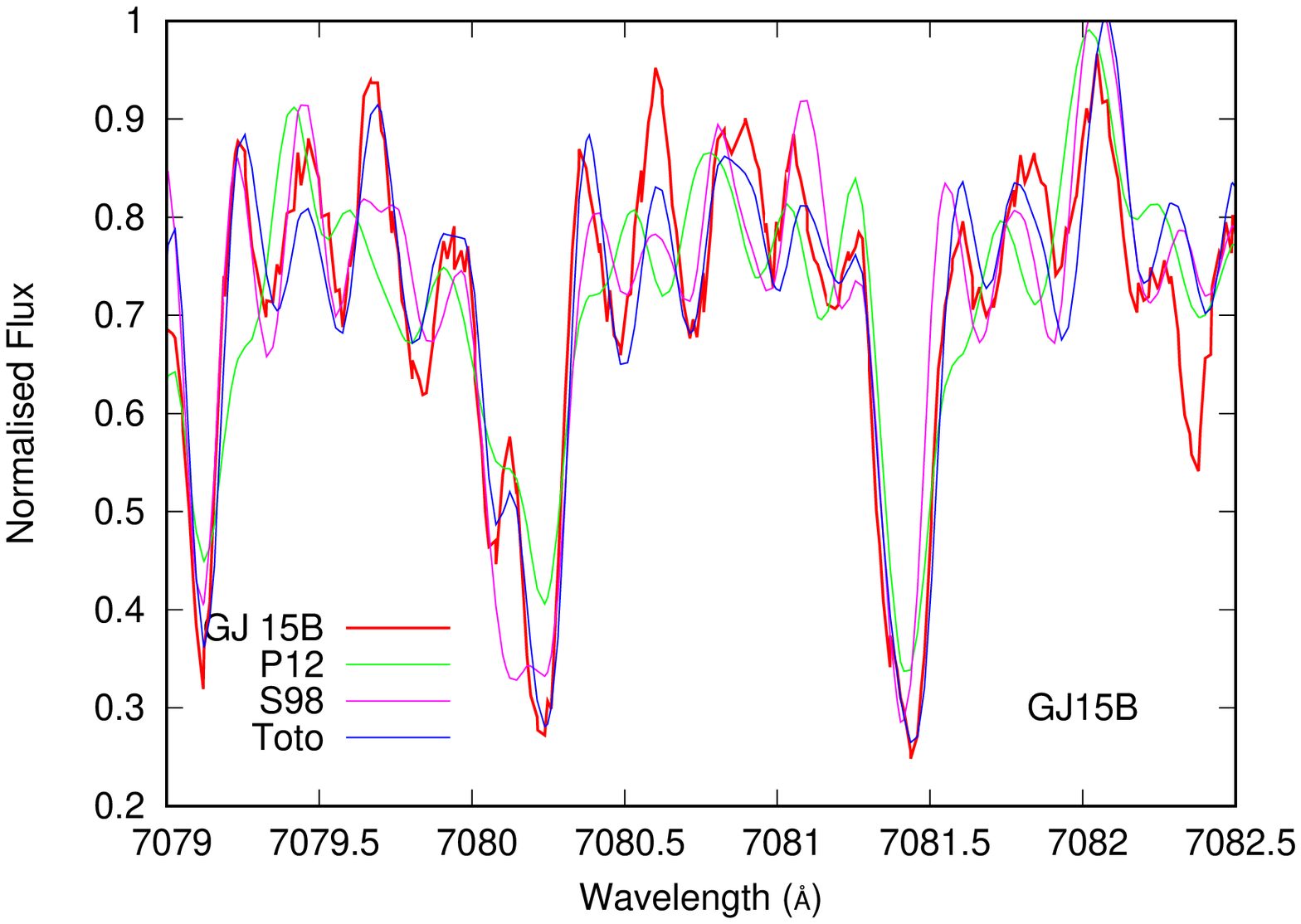}

\caption{\label{_cAB} Comparison of fits to the observed TiO features across $x_1$ spectral range in GJ 15A {\it (left panels)} and
GJ 15B {\it (right panels)} using line lists of different authors. Here solar isotopic ratios of Ti was adopted.
}
\end{figure*}

\begin{figure*}
   \centering
   \includegraphics[width=0.98\columnwidth]{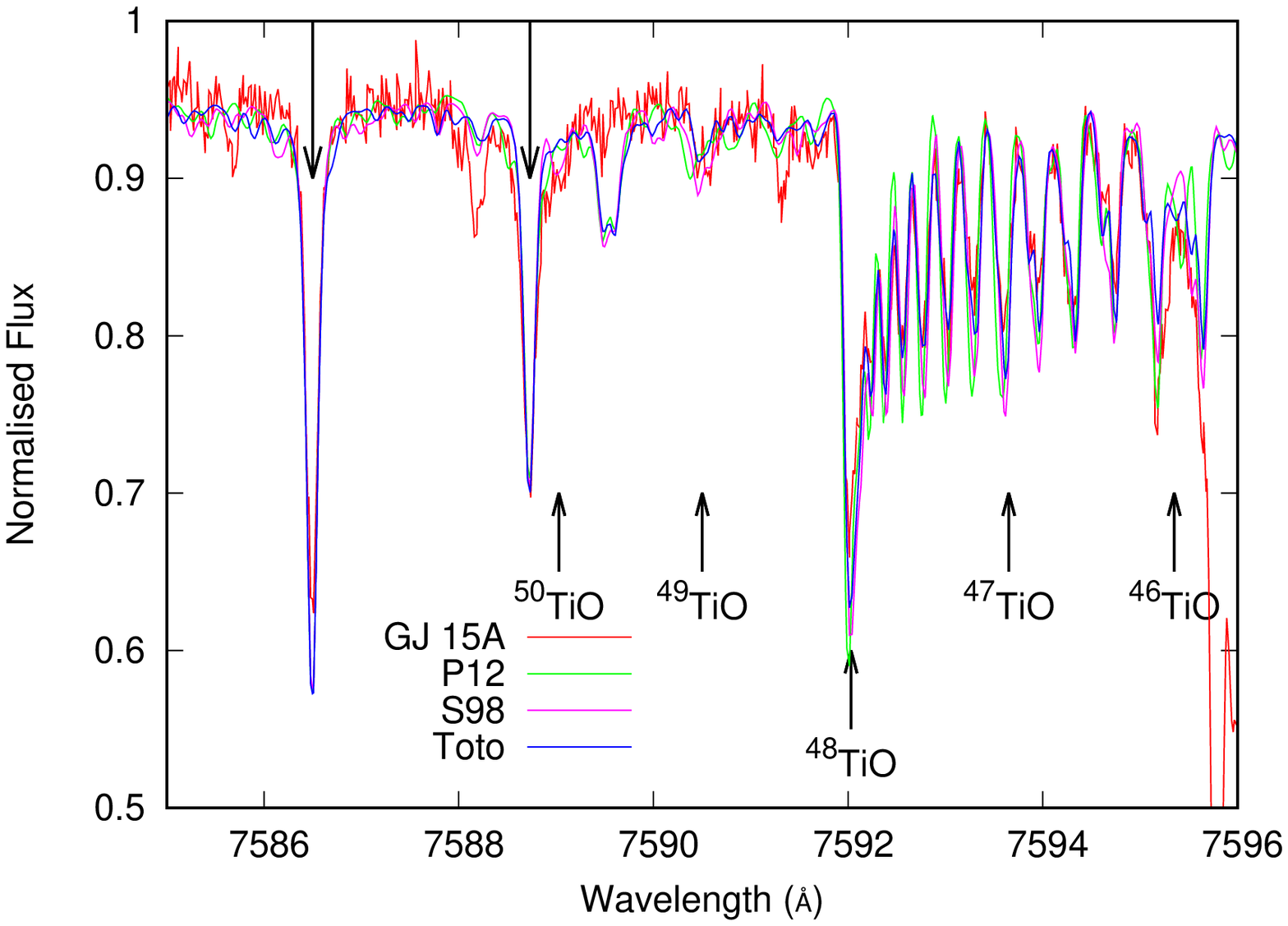}
    \includegraphics[width=0.98\columnwidth]{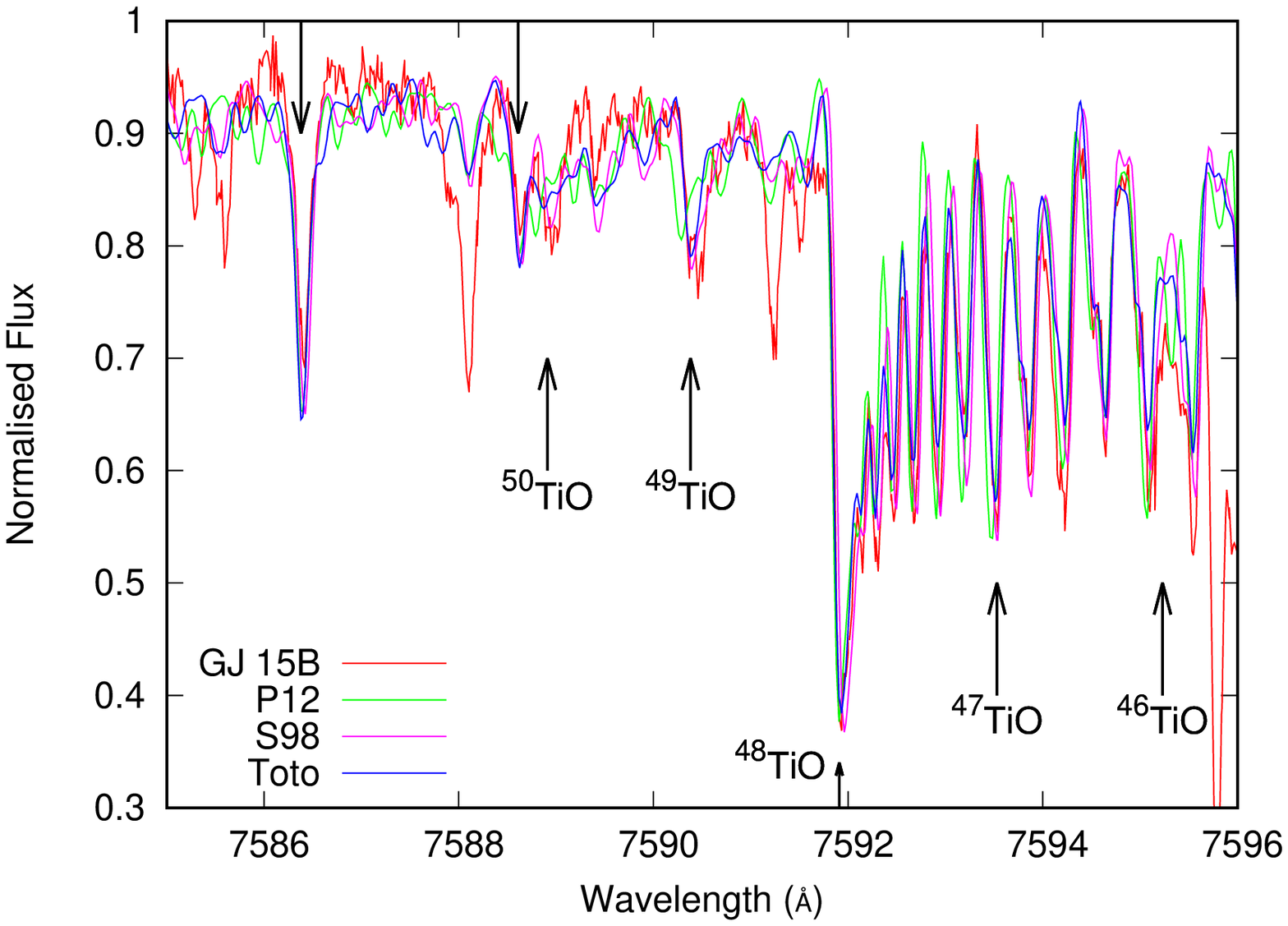}
\caption{\label{_xxc} Comparison of fits to the observed TiO features across $x_2$ spectral range in GJ 15A {\it (left panel)} and
GJ 15B {\it (right panel)} using line lists of different authors. Here solar isotopic ratios of Ti was adopted.
}
\end{figure*}

\begin{figure*}
   \centering
   \includegraphics[width=0.98\columnwidth]{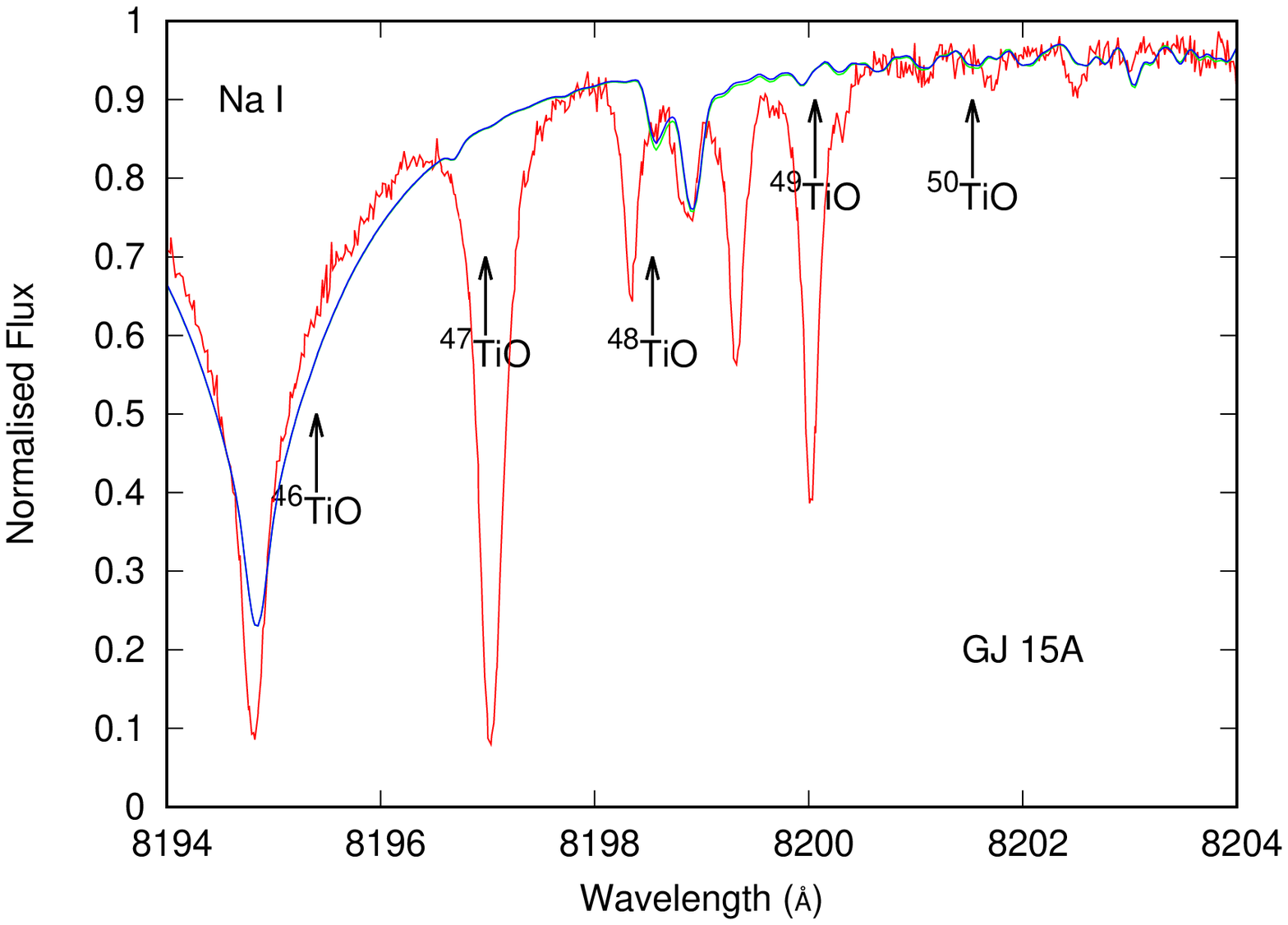}
    \includegraphics[width=0.98\columnwidth]{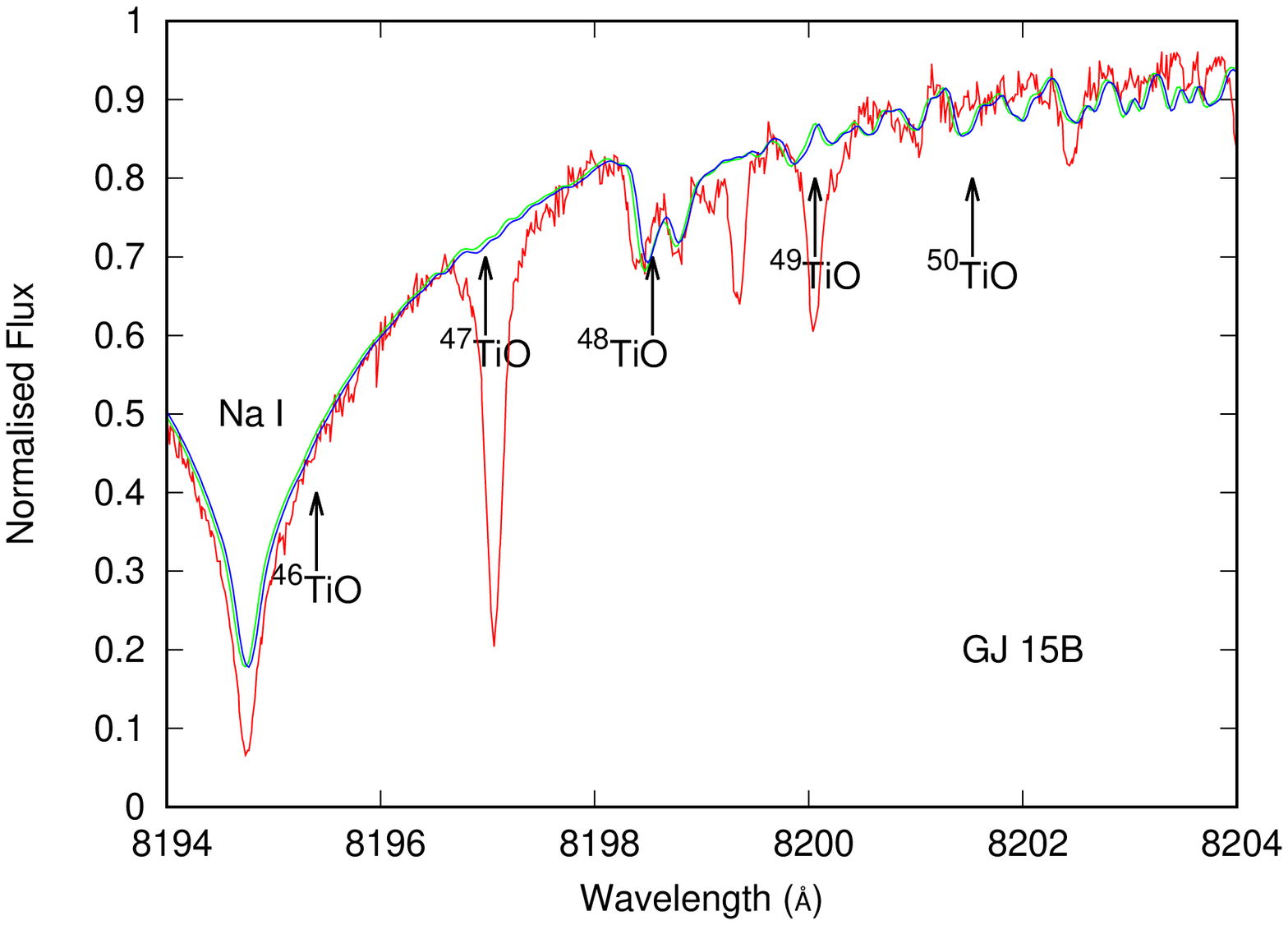}
    \includegraphics[width=0.98\columnwidth]{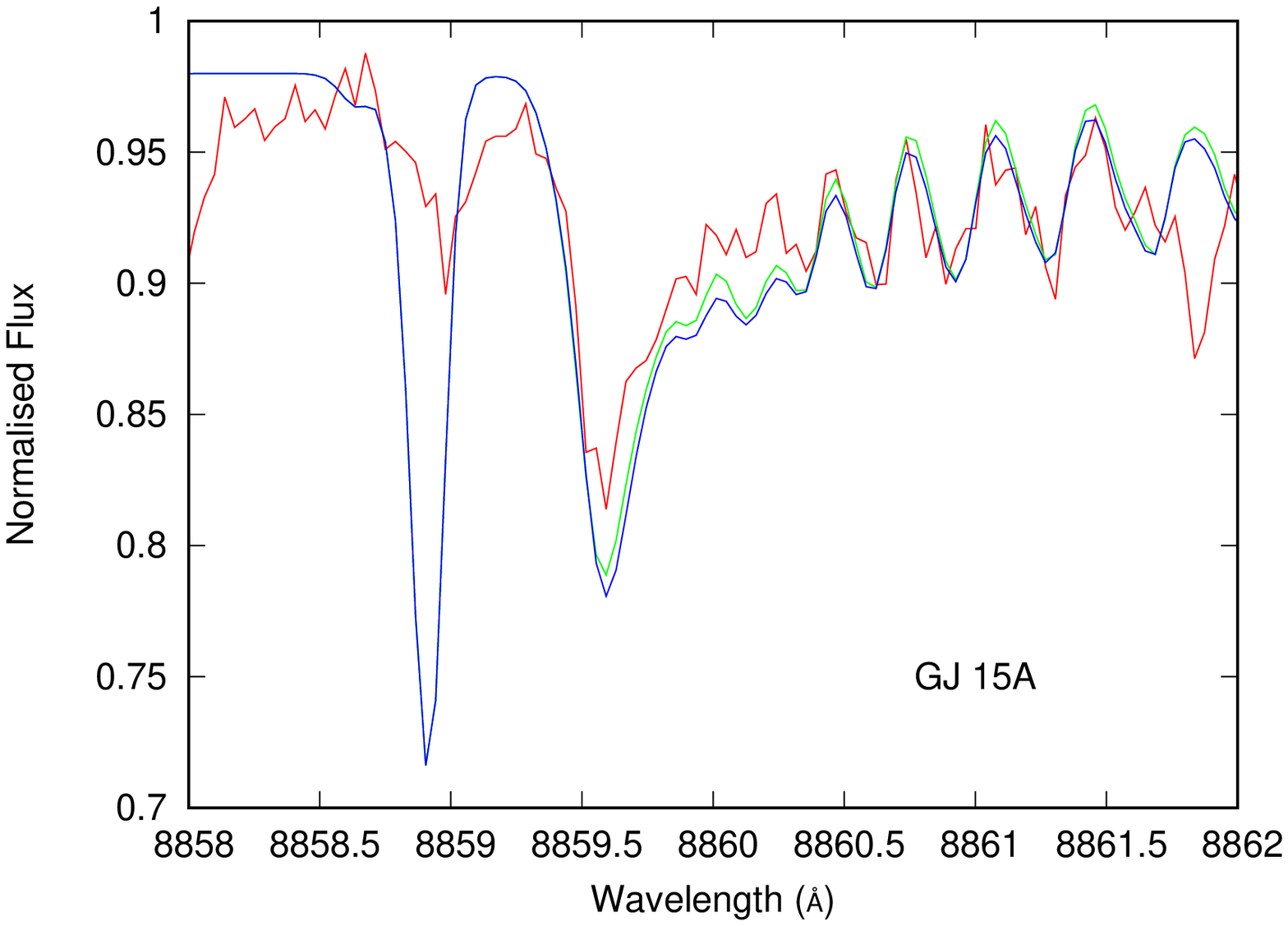}
    \includegraphics[width=0.98\columnwidth]{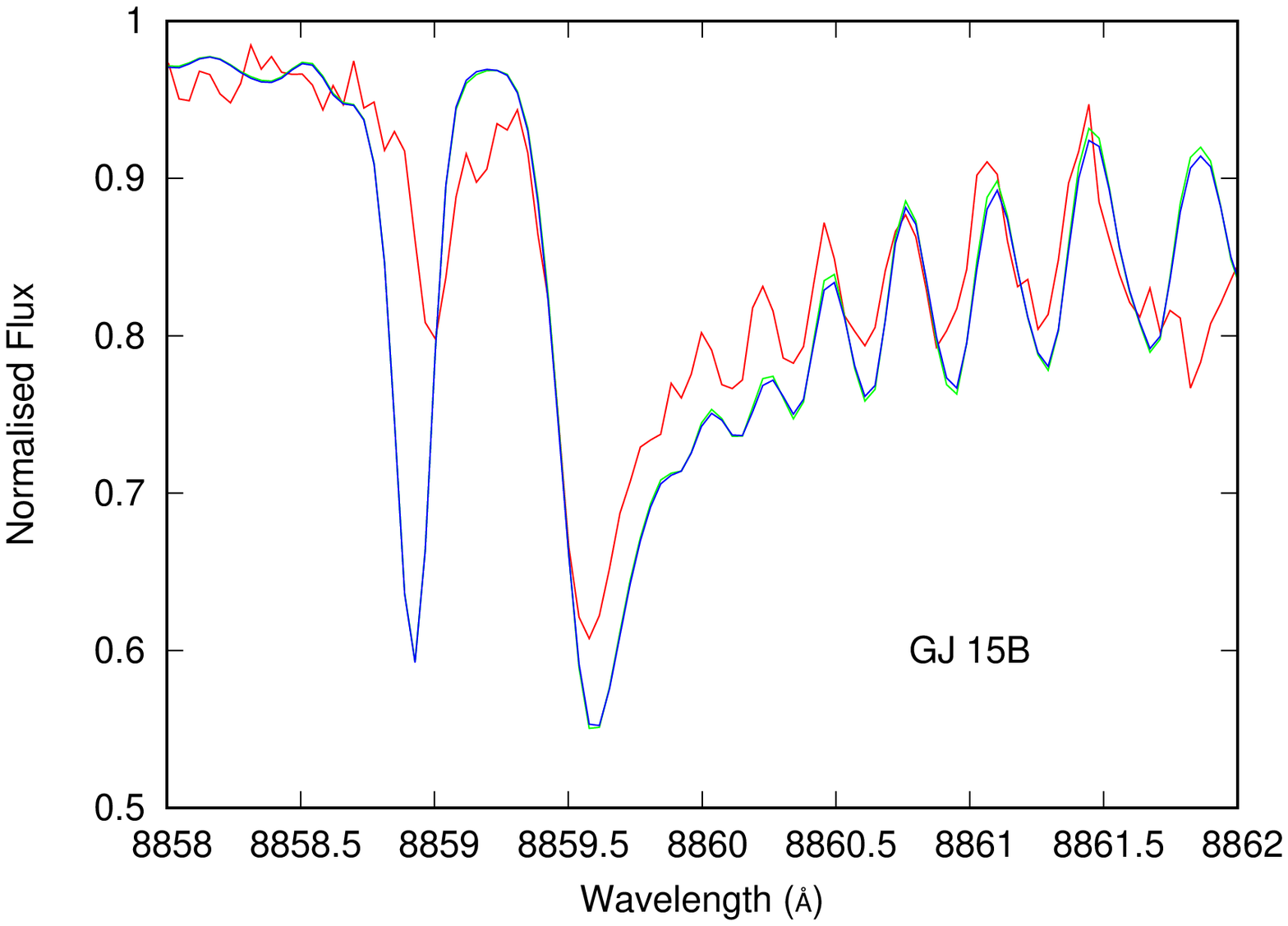}
    \includegraphics[width=0.98\columnwidth]{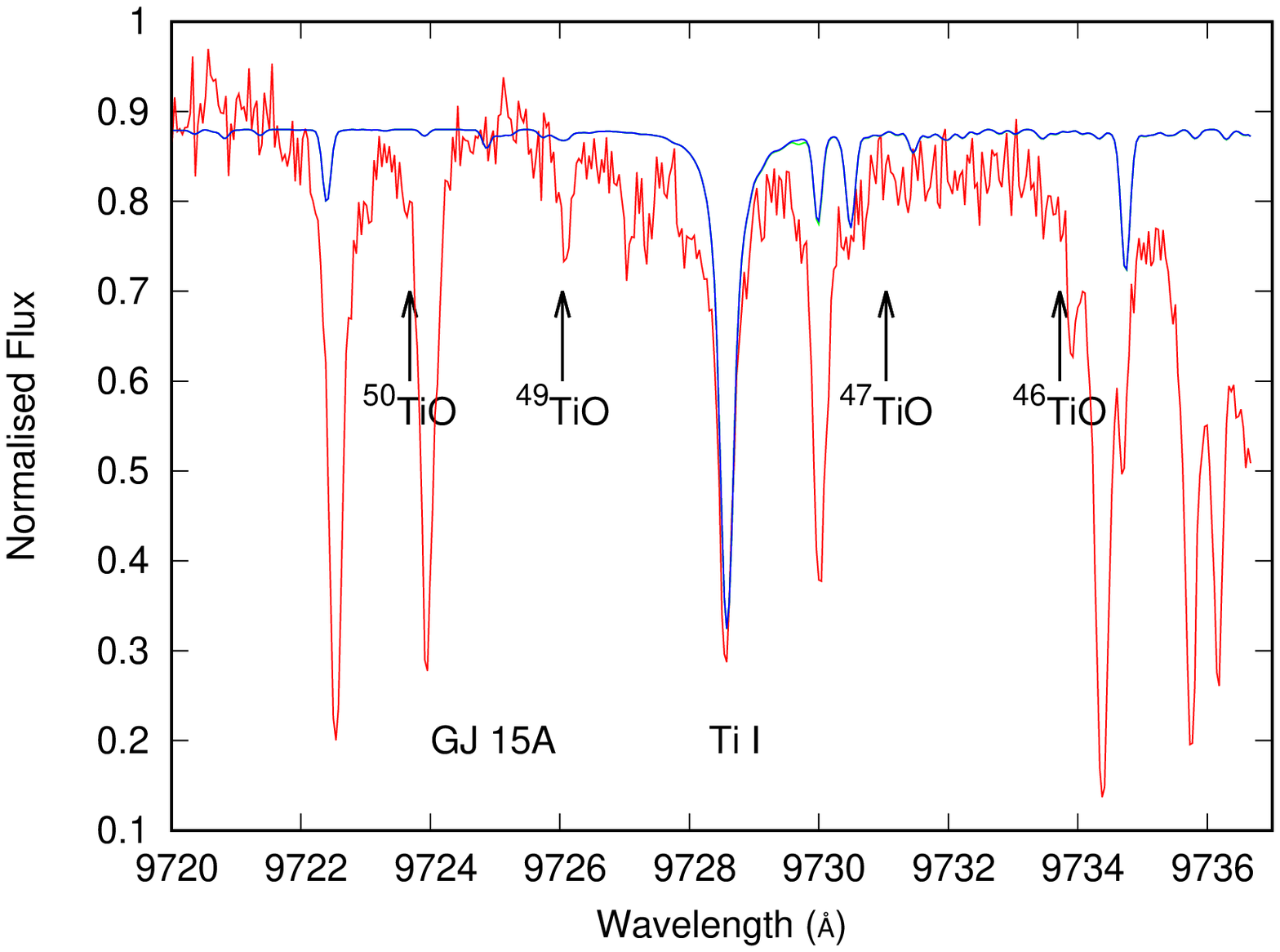}
    \includegraphics[width=0.98\columnwidth]{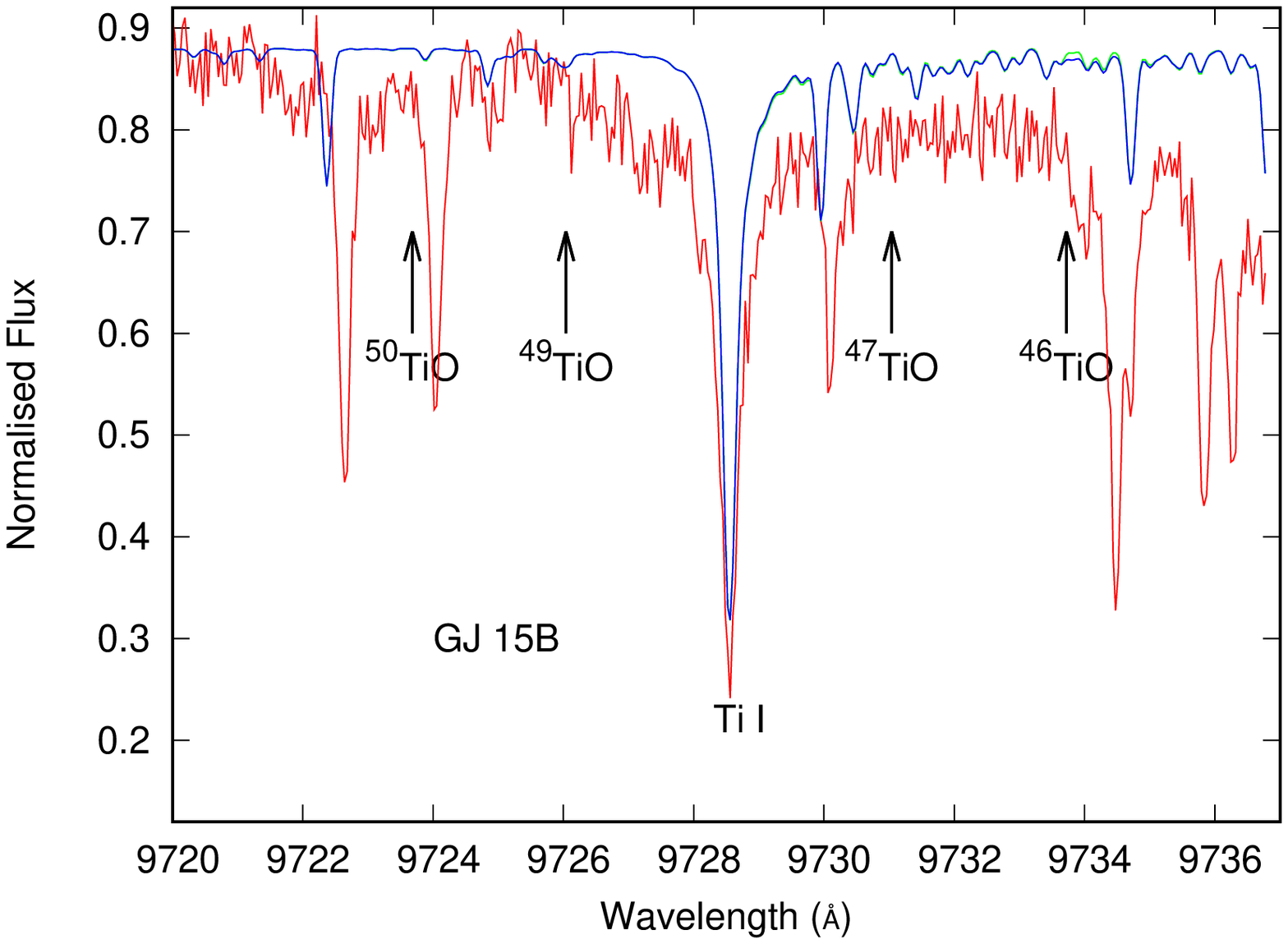}

\caption{\label{_AAB} {\it Top left:} Comparison of observed and computed
TiO isotopic ratios spectra of GJ 15A and GJ 15B across $x_3$ {\it (top panels)},
$x_4$ {\it (middle panels)} and $x_5$ {\it (bottom panels}) spectral ranges for the solar (green line) and M25 (blue line) abundance ratios.
}
\end{figure*}

\end{document}